\documentclass[11pt]{article}

\usepackage[margin=1in]{geometry}
\usepackage[font={small,it}]{caption}

\usepackage{amsmath,amssymb,amsthm,mathtools}
\usepackage{thm-restate}
\usepackage{algorithm}
\usepackage{parskip}
\usepackage{tikz,nicefrac}
\usepackage{bbm}
\usepackage[dvipsnames]{xcolor}
\usetikzlibrary{arrows.meta,decorations.pathreplacing}
\usepackage{hyperref}
\usepackage{cleveref}
\hypersetup{
    hypertexnames=false,
    colorlinks=true,  
    linkcolor=blue,   
    citecolor=blue,  
    urlcolor=blue,    
}
\usepackage[normalem]{ulem}
\usepackage{thmtools}

\newtheorem{theorem}{Theorem}[section]
\newtheorem*{theorem*}{Theorem}
\newtheorem{lemma}[theorem]{Lemma}
\newtheorem{proposition}[theorem]{Proposition}
\newtheorem{corollary}[theorem]{Corollary}
\newtheorem{claim}[theorem]{Claim}
\newtheorem{remark}[theorem]{Remark}

\newcommand{\E}[1]{\mathbb E\left[#1\right]}
\newcommand{\Esub}[2]{\mathbb E_{#1}\left[#2\right]}
\renewcommand{\P}[1]{\mathbb P\left(#1\right)}
\newcommand{\R}{\mathbb R}
\newcommand{\Rbb}{\mathbb R}
\newcommand{\Et}{\mathbb E_t}
\DeclareMathOperator{\Cov}{Cov}
\DeclareMathOperator{\sd}{sd}
\DeclareMathOperator{\Tr}{Tr}

\newcommand{\norm}[1]{\lVert #1\rVert_2}

\newcommand{\bmu}{\boldsymbol\mu}
\newcommand{\bm}{\boldsymbol m}
\newcommand{\bs}{\boldsymbol s}
\newcommand{\bb}{\boldsymbol b}
\newcommand{\bx}{\boldsymbol x}
\newcommand{\bz}{\boldsymbol z}

\title{Contextual Search and Bilateral Trade\\ with a Trace-Square-Root Potential}
\author{Romain Cosson\thanks{Computer Science Department, New York University, NYC, United States.}
\and Federico Fusco\thanks{Department of Computer, Control, and Management Engineering, Sapienza University, Rome, Italy.}
\and Anupam Gupta\footnotemark[1] \footnotemark[3]
\and Stefano Leonardi\footnotemark[2]
\and Renato Paes Leme\thanks{Google Research, NYC, United States.}
\and Matteo Russo\thanks{EPFL, Lausanne, Switzerland.}}
\date{}

\begin{document}
\maketitle
\begin{abstract}
We present a unified potential method for contextual search and bilateral trade with one-bit feedback and obtain order-tight algorithms for both problems.
Our framework starts from a Bayesian perspective in which the unknown target (for contextual search) and the unknown pair of valuations (for bilateral trade) are sampled from a known prior distribution.
In that model, the decision maker bases its prediction on the posterior distribution, obtained by conditioning on feedback received in previous rounds. 
For Bayesian contextual search, we present an algorithm with a {total loss bounded by} $O(\Tr \sqrt \Sigma)${, where} $\Sigma$ is the covariance matrix of the target's prior; 
and for contextual bilateral trade, our algorithm's regret is bounded by $O(\Tr \sqrt{\Sigma_S}+\Tr \sqrt{\Sigma_B})$, where $\Sigma_S$ and $\Sigma_B$ are the covariance matrices of the seller's and buyer's valuations. 

We then use ideas from this Bayesian setting and adapt a ``dual'' of this trace-square-root potential to the classical worst-case setting, where the target and valuations are chosen adversarially from the unit Euclidean ball. For both contextual search and bilateral trade, we obtain tight $O(\sqrt d)$-regret algorithms. These results match the existing lower bounds up to a constant {factor and improve on} the previous guarantees of $O(d\log d)$ for contextual {search and} of $O(d6^d)$ for bilateral trade. 
\end{abstract}
\thispagestyle{empty}

\newpage
\pagenumbering{arabic}

\section{Introduction}\label{sec:intro}

We study parameter search problems in settings where decision makers
receive only limited feedback. For example, a pricing algorithm sees
the features of the customer, quotes a price for a product described
by a feature vector and learns only whether the customer bought it. An
idealized dosing procedure receives {patient features and learns} only
whether the dose was too low or too high. A broker quotes prices to
both a buyer and a seller, and learns only whether they traded. In
each case, there is a vector hidden from the learner, and {at each step,
the observed features provide a ``context'': together, the hidden vector
and the context determine a target value that the learner tries to match.} In this paper, we study two classical
problems at the intersection of theoretical computer science and
economics that are inspired by such situations: \textit{contextual
  search} and \textit{contextual bilateral trade}.

\textbf{Contextual search.}  The basic problem in this setting is
\emph{contextual search}, where the target is an unknown vector
$\bz\in\R^d$. In round $t$, an adversary reveals the \emph{context}
vector $\bx_t$ with $\norm{\bx_t}\le1$. The two together define the
target value $\bx_t^\top \bz$---this is unknown to the learner, since
$\bz$ {is unknown}. The learner does observe $\bx_t$, then proposes a
price $p_t$ and incurs symmetric loss
$\ell_t=|\bx_t^\top\bz-p_t|$. The learner {observes neither the target
value nor the loss incurred}, but only the one-bit feedback
$\operatorname{sign}(\bx_t^\top\bz-p_t)$---i.e., whether the price
was too high or too low. Using this information, the learner aims to
bound {its} total loss $\sum_{t=1}^T\E{\ell_t}$ (also called the
``regret'') independently of $T$. The problem of contextual search
arose from feature-based pricing and auctions
\cite{KleinbergL03,CLPL16,LPLV18}. It is useful to {note} that, for $d=1$, the
problem reduces to binary search, but higher-dimensional versions
become much more
challenging. 
Indeed, a series of works {has studied} this problem, and the
{state of the art} is an $O(d \log d)$-competitive
algorithm~\cite{LPS21}. However, the lower bound was only
$\Omega(\sqrt{d})$: can we close this gap?

\textbf{Contextual bilateral trade.}  An extension of contextual
search is \emph{bilateral trade}, where the learner is a broker who
intermediates between a seller and a buyer with hidden vectors
$\boldsymbol s,\boldsymbol b\in\R^d$.  In round $t$, an adversary
reveals a context vector $\bx_t$ with $\norm{\bx_t}\le1$ corresponding to valuations
$\bx_t^\top \boldsymbol s$ for the seller and $\bx_t^\top \boldsymbol b$ for the buyer---{these} are unknown to the broker, {who} does not see $\boldsymbol s, \boldsymbol b$. The broker then proposes a price $p_t$ and observes whether both agents are willing to trade at that price, i.e., if
$\bx_t^\top \boldsymbol s \leq p_t \leq \bx_t^\top \boldsymbol b$. {Note that, given the feedback structure,} a failed trade does not identify which agent refused. The
broker aims to maximize gain from trade, defined as
$\sum (\bx_t^\top \boldsymbol b-\bx_t^\top \boldsymbol s)_+$, where the sum
is taken over all rounds $t$ in which the trade took place, or
equivalently, to minimize the total loss $\sum_{t=1}^T\E{\ell_t}$, which is the total gain
from trade it failed to realize. The model was introduced by Gaucher
et al.~\cite{GBCCP25}, building on the online bilateral trade
framework of Cesa-Bianchi et
al.~\cite{CCCFL21,Cesa-BianchiCCF24}. 
The gaps left by the literature are much larger in this setting: the only upper bound for
this one-bit feedback model {is} $O(d 6^d)$~\cite{CFGLPR26}, and the
lower bound is again $\Omega(\sqrt{d})$, leaving a huge gap: can we
{close it?}

\subsection{Results}

In this work, we give tight results for both these problems: we show
that both problems admit $O(\sqrt{d})$-regret algorithms, which are
tight {by the lower bounds mentioned above}. In addition to giving tight
results, however, our work showcases the use of a potential
function---the \emph{square-root trace potential}---which is the
{central tool in both analyses}. Moreover, we also present results
for the Bayesian setting (where $\bz$, $\boldsymbol s$, and $\boldsymbol b$ are drawn from known prior
distributions), which inspire our algorithms for the adversarial
setting.

\paragraph{Worst-case guarantees.} For worst-case guarantees, the
unknown target $\bz\in \Rbb^d$ and the valuation vectors
$\boldsymbol s,\boldsymbol b\in\R^d$ are assumed to be selected from
the Euclidean unit ball by the adversary. The adversary knows the
learner's algorithm, but not the random bits it uses. This is the main
point of view taken in prior works on both contextual search and
bilateral trade. We obtain horizon-independent $O(\sqrt d)$
expected-loss bounds for both problems, matching the existing lower
bounds up to constant factors. Table~\ref{tab:results} summarizes our
worst-case guarantees and compares them with {prior work} for
contextual search and bilateral
trade. 

\begin{table}[htbp]
\centering
\renewcommand{\arraystretch}{1.2}
\begin{tabular}{lll}
\hline
Result & Contextual search & Bilateral trade \\
\hline
Standard lower bound \cite{LPS21,CFGLPR26}& $\sqrt d$  & ${\sqrt d/6}$ \\
Paes Leme and Schneider \cite{PLS18} & $O(d^4)$ & --- \\
Liu, Paes Leme, and Schneider \cite{LPS21} & $O(d\log d)$ & --- \\
Cosson et al.~\cite{CFGLPR26} & --- & $O(d\,6^d)$ \\
\textbf{This paper} (Thms.~\ref{thm:main} and~\ref{thm:trade-direct}) & $O(\sqrt d)$ (Alg.~\ref{alg:adversarial})& $O(\sqrt d)$ (Alg.~\ref{alg:trade-adversarial}) \\
\hline
\end{tabular}
\caption{Worst-case guarantees for unit-norm contexts and hidden vectors.}
\label{tab:results}
\end{table}

\paragraph{Bayesian guarantees.} For Bayesian guarantees, we assume that the unknown vectors $\bz, \bs, \bb \in \Rbb^d$ are sampled from a known prior distribution with finite second moments. This point of view deviates from the existing literature but is particularly illuminating: it yields simpler algorithms and a first potential function, based on posterior covariance, that will guide the adversarial analysis. For known priors with finite second moments, our bounds depend only on the prior covariance matrices and hold at every horizon (Table~\ref{tab:bayesian-results}).

\begin{table}[htbp]
\centering
\setlength{\tabcolsep}{10pt}
\renewcommand{\arraystretch}{1.35}
\begin{tabular}{lll}
\hline
Result & Contextual search & Bilateral trade \\
\hline
\textbf{This paper} (Thms.~\ref{thm:bayesian-contextual-search} and~\ref{thm:trade-bayes}) & $O(\Tr\sqrt{\Sigma})$ (Alg.~\ref{alg:bayesian})& $O(\Tr\sqrt{\Sigma_S}+\Tr\sqrt{\Sigma_B})$ (Alg.~\ref{alg:trade-bayesian}) \\
\hline
\end{tabular}
\caption{Bayesian rates for unit-norm contexts and priors with finite second moments. $\Sigma$ is {the prior covariance in contextual search}, and $\Sigma_S,\Sigma_B$ are the seller's and buyer's marginal prior covariances in bilateral trade. }
\label{tab:bayesian-results}
\end{table}

Our worst-case guarantees and our Bayesian guarantees are clearly
related, since $\Tr\sqrt{\Sigma}= O(\sqrt d)$ when $\Sigma$ is the
covariance matrix {associated with} the uniform distribution in the
Euclidean ball. Yet neither directly implies the other: in particular,
the Bayesian setting allows for distributions with unbounded support
or with skewed covariance matrices, such as Gaussian
distributions. Thus, the Bayesian setting is not merely an
{intermediate} conceptual step toward worst-case competitive guarantees,
but is of independent interest.

\subsection{Our Techniques}

{We now outline the intuition behind our algorithms and the essential
ingredients of their analysis.}

\subsubsection{Our Algorithms}

In the Bayesian setting, the learner is given a prior distribution on
the hidden {parameters}: the vector $\bz$ for contextual search, or
the pair $(\bb, \bs)$ for contextual bilateral trade. Using this, at
any time $t$, the learner has access to the \emph{posterior}
distribution of the unknown parameters $\bz$ or $\bs,\bb$ conditioned
on all the past observations. Our Bayesian algorithms
(\Cref{alg:bayesian,alg:trade-bayesian}) compute a proposed price
$p_t$ based only on that posterior distribution and on the context
vector $\bx_t$; then they update the posterior distribution with the corresponding feedback. Our Bayesian
algorithm for contextual search (\Cref{alg:bayesian}) is remarkably
simple: it posts prices $\mu_t \pm \nicefrac{\sigma_t}{2}$, where
$\mu_t$ is the posterior mean of $\bx_t^\top\bz$, and $\sigma_t$ is the
posterior standard deviation of $\bx_t^\top\bz$. 
{The bilateral trade algorithm (\Cref{alg:trade-bayesian}) similarly uses the posterior, now of the two projected values $\bx_t^\top\bs$ and $\bx_t^\top\bb$, and posts a common price to both agents. This price is chosen to balance trading and learning, allowing us to bound total regret by the learner's initial uncertainty (Lemma~\ref{lem:toi}).}

In the worst-case setting,
the learner has no
posterior distribution on the unknown parameters, {so it maintains a suitable ``proxy'' {for the potential} used in the Bayesian analysis}. For contextual search, it
maintains two parameters, a vector $\bm_t\in \Rbb^d$ and a matrix
$R_t\in \Rbb^{d\times d}$, which we call the posterior \textit{mean
  proxy} and the posterior \textit{square-root covariance
  proxy}. Informally, these parameters can be interpreted as an
estimate of the unknown parameter (for $\bm_t$) and a measure of
uncertainty in that estimate in all directions (for
$R_t$). Importantly, we do not maintain explicit uncertainty sets as
in previous work---in particular, the unknown parameter need not lie
within the ellipsoid $\bm_t + R_tB_d$, where $B_d$ is the unit
Euclidean ball (as, e.g., in the {ellipsoid} method). Instead, our
proofs directly show that the per-step regret can be charged to the
{decrease in} a certain potential. {From this perspective, the worst-case
contextual search algorithm (\Cref{alg:adversarial})} posts prices
$m_t \pm \nicefrac{r_t}{2}$, where $m_t = \bx_t^\top\bm_t$, and
$r_t = \norm{R_t \bx_t}$---{a rule remarkably similar to the Bayesian rule in}
\Cref{alg:bayesian}.
{For bilateral trade (\Cref{alg:trade-adversarial}), $\bm_t\in\Rbb^{2d}$ stores separate $d$-dimensional mean proxies for the seller and buyer, and the matrix $R_t\in\Rbb^{2d\times2d}$ is block diagonal, with one $d\times d$ block for each agent's square-root covariance proxy. The algorithm samples a common price and two possible successor states from a finite lottery, then keeps the successor selected by the trade bit. {Whereas contextual search admits explicit state updates, here we apply} the minimax theorem to guarantee the existence of a single lottery that works for every {possible} hidden pair of buyer and seller.}

Although the focus of the paper is not on computational complexity, we
note that {our worst-case contextual search algorithm (\Cref{alg:adversarial})} uses
$O(d^2)$ memory throughout the execution and performs only $O(d^2)$
operations per round, making it significantly {more practical than}
previous approaches for contextual search that rely on costly
geometric routines such as computing intrinsic volumes or barycenters
of high-dimensional convex bodies \cite{PLS18,LPS21}.

\subsubsection{Analyses via the Trace-Sqrt Potential}\label{sec:potential}

\newcommand{\agsqrt}[1]{(#1)^{1/2}}

\begin{figure}[t]
\centering
\begin{tikzpicture}[x=1.35cm,y=1.35cm,>=Stealth,line cap=round,line join=round]
 \definecolor{regionblue}{RGB}{54,95,136}
 \definecolor{cutorange}{RGB}{180,85,35}
 \begin{scope}
  \node at (0,1.6) {(a) Principal axes};
  \path[fill=regionblue!15,draw=regionblue,line width=1pt]
    (0,0) ellipse[x radius=1.4,y radius=.7];
  \draw[->,black!55,line width=.65pt] (-1.65,0)--(1.8,0)
    node[right,inner sep=3pt,text=black] {$\boldsymbol e_1$};
  \draw[->,black!55,line width=.65pt] (0,-.95)--(0,1.08)
    node[above,inner sep=3pt,text=black] {$\boldsymbol e_2$};
  \draw[->,cutorange,line width=.9pt] (0,0)--(1.4,0)
    node[midway,above,inner sep=3pt] {$\sigma_1$};
  \draw[->,cutorange,line width=.9pt] (0,0)--(0,.7)
    node[midway,left,inner sep=3pt] {$\sigma_2$};
  \fill[regionblue] (0,0) circle[radius=2.2pt];
  \node[anchor=north west,inner sep=3pt,text=regionblue]
    at (0,0) {$\bmu$};
 \end{scope}
 \begin{scope}[xshift=6.6cm]
  \node at (0,1.6) {(b) Enclosing box};
  \path[fill=regionblue!15,draw=regionblue,line width=1pt]
    (0,0) ellipse[x radius=1.4,y radius=.7];
  \draw[cutorange,line width=.8pt] (-1.4,-.7) rectangle (1.4,.7);
  \fill[regionblue] (0,0) circle[radius=2.2pt];
  \node[anchor=north west,inner sep=3pt,text=regionblue]
    at (0,0) {$\bmu$};
  \draw[black!35,line width=.5pt] (-1.4,-.77)--(-1.4,-1.2);
  \draw[black!35,line width=.5pt] (1.4,-.77)--(1.4,-1.2);
  \draw[<->,cutorange,line width=.8pt] (-1.4,-1.08)--(1.4,-1.08)
    node[midway,below,inner sep=3pt] {$2\sigma_1$};
  \draw[black!35,line width=.5pt] (1.47,-.7)--(1.95,-.7);
  \draw[black!35,line width=.5pt] (1.47,.7)--(1.95,.7);
  \draw[<->,cutorange,line width=.8pt] (1.83,-.7)--(1.83,.7)
    node[midway,right,inner sep=3pt] {$2\sigma_2$};
 \end{scope}
\end{tikzpicture}
\caption{The covariance ellipsoid associated with $\Sigma$. Geometrically, the trace-square-root potential $\Psi = \sigma_1 + \sigma_2$ equals the $1$-dimensional intrinsic volume (i.e., {half the perimeter}) of the smallest enclosing box of the ellipsoid, divided by $2$. This perspective relates our potential to {early geometric approaches to} contextual search \cite{PLS18}{, which relied} on intrinsic volumes. }
\label{fig:covariance-ellipse}
\end{figure}
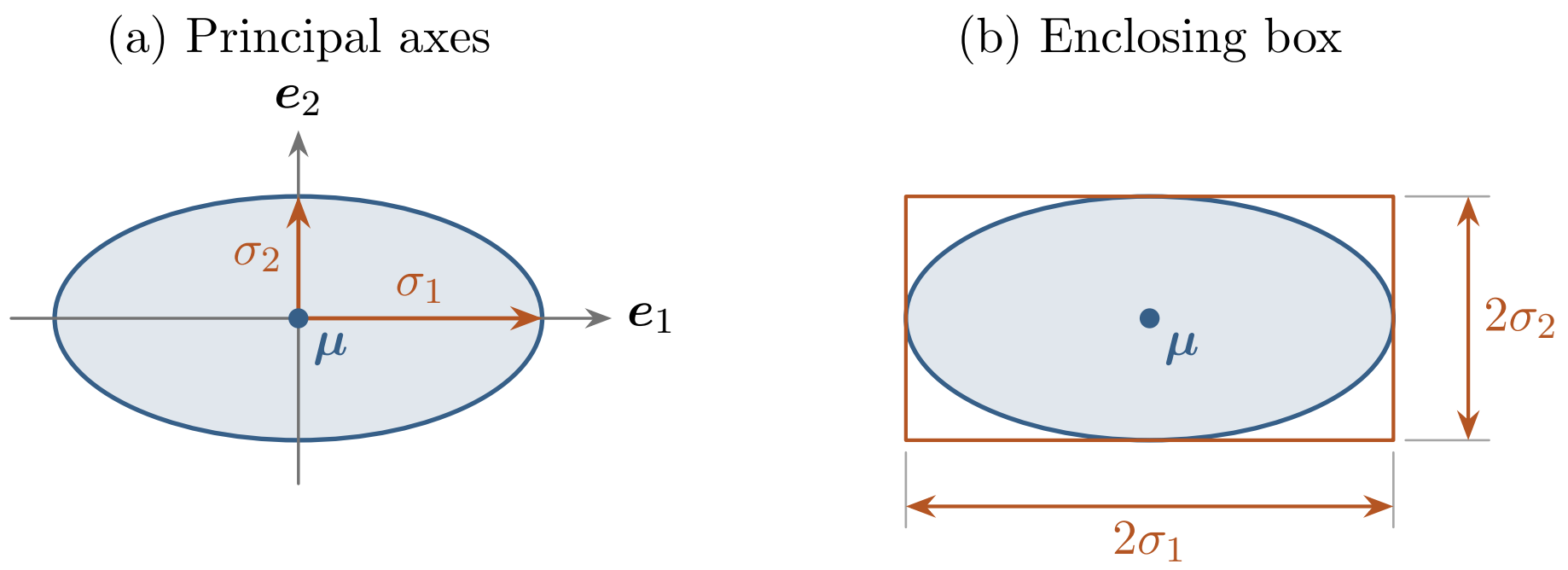

We bound the regret of all our algorithms via a potential-based
analysis, using a pair of potential functions we feel may be useful in
other contexts.  It will be illuminating to start with the Bayesian
problem {for contextual search}. Here we have a random vector $\bz\in\R^d$ with finite second
moment; this represents the hidden vector that a learner does not
know; however, we do know the distribution {from} which $\bz$ is drawn,
and hence know its mean $\bmu=\E{\bz}${\ and covariance}
$\Sigma=\Cov(\bz)\succeq0$. In order to quantify the uncertainty about
$\bz$, we use the \emph{trace-sqrt potential} of the covariance:
\begin{equation}\label{eq:trace-sqrt-potential}
 \Psi(\Sigma)=\Tr\sqrt\Sigma = \sum_{i=1}^d \sigma_i,
\end{equation}
where $\sigma_i = \sqrt{\lambda_i}$ is the square root of the $i$-th
eigenvalue of $\Sigma$, commonly called the $i$-th semiaxis {length; see} \Cref{fig:covariance-ellipse} for an illustration. {The figure highlights a key point:} the
potential naturally captures a notion of ``distance'' and ``spread'',
as opposed to other potentials (like the logarithm of the volume)
which would need various tricks to relate regret to the
reduction in the volume, incurring losses along the way.

Whatever the learner observes next --- the {feedback from a round}, a new
{context, or any other information} revealed about $\bz$ --- is a
\emph{signal}: a random variable $Y$ defined jointly with $\bz$. After
observing $Y$, the remaining uncertainty is described by $\bz\mid Y$;
let the new covariance matrix be $\Sigma_Y$. This means the potential
moves from $\Psi(\Sigma)$ to $\Psi(\Sigma_Y)$.

\paragraph{The effect of signals on the potential.}
The first important observation is this: For any signal $Y$, the law
of total covariance ({which states that} the total covariance equals the expected posterior covariance plus the covariance of the posterior mean) gives
\begin{gather}
  \Cov(\bz) \;\succeq\;\E{\Cov(\bz\mid Y)}. \label{eq:total-cov}
\end{gather}
Now, the monotonicity and concavity of the operator $A \mapsto \sqrt{A}$ on the positive-semidefinite cone (see, e.g., Example 3.48 in \cite{vandenberghe2004convex}) allow us to infer that the potential is
monotone in the L\"owner order. This, combined with
(\ref{eq:total-cov}){, means that} 
\begin{equation}\label{eq:potential-drops}
    \Psi(\Sigma) \geq \E{\Psi(\Sigma_Y)}
\end{equation}
and hence the potential is non-increasing in expectation. Of course, this
expected decrease is not good enough---{we need the
potential decrease to pay for our regret}.

\paragraph{Potential drops versus directional uncertainty.}
The {key ingredient in relating} the potential drop to our regret is
the following claim, which is a strengthening of the matrix
square-root function's {monotonicity and concavity}: 

\begin{lemma}[Gap Function]
  \label{lem:g-mono-concave}
  For any vector
  $\bx\in \Rbb^d$ with $||\bx||\leq 1$, the ``gap'' function
  \[ g_{\bx}(A) = \Tr\sqrt A - \sqrt{\bx^\top A \bx}
  \] is monotone
  and concave on the positive-semidefinite cone.
\end{lemma}

We defer the proof to
\Cref{lem:envelope} in Appendix~\ref{app:gap-concavity}, but let us just
mention that it follows by writing $g_{\bx}$ as the infimum of a
family of affine functions{:}
\begin{gather}
  2g_{\bx}(A)=\inf_{R\succ0}\left\{
    \Tr R-\norm{R\bx}
    +\Tr\!\left[\left(R^{-1}-\frac{\bx\bx^\top}{\norm{R\bx}}\right)A\right]\right\}, \label{eq:affine}
\end{gather}
where $\bx\bx^\top/\norm{R\bx} = 0$ when $\bx = 0$. (Somewhat
surprisingly, we will return to this expression in a moment!)
Now the potential drop can be {bounded below by the decrease in} the directional standard deviation, using (\ref{eq:total-cov}), \Cref{lem:g-mono-concave}, and Jensen's inequality:

\begin{lemma}[Potential decrease lemma]\label{lem:directional-sd-decrease}
  Let $\bz\in\R^d$ be a random vector with {finite second moment and covariance matrix}
  $\Sigma \succeq 0$, and let $Y$ be any jointly defined random
  variable. Fix $\bx\in\R^d$ with $\norm{\bx}\le1$. Then
\begin{gather}
  \underbrace{\Tr(\sqrt \Sigma) - {\E{\Tr (\sqrt
    \Sigma_Y)}}}_{\text{expected potential drop}} \geq
  \underbrace{\sqrt{\bx^\top\Sigma\bx}
      -\E{\sqrt{\bx^\top\Sigma_Y\bx}}}_{\text{drop in directional std.\ dev.}}.  \label{eq:blah}
\end{gather}
\end{lemma}

To get intuition for the quantity on the {right-hand side}, define the ``uncertainty'' along a specific direction $\bx$ (which should be thought of as the context vector) by the ``directional'' standard deviation
$\sd_{\bx}(\bz)=\sqrt{\bx^\top\Sigma\bx}$.
So the {right-hand side} of \eqref{eq:blah} is indeed the drop in the standard deviation in the direction of the context vector. 
In a very concrete sense, this reduces the high-dimensional object
(related to $\Sigma$) to a one-dimensional object (the reduction in
standard deviation of a scalar random variable).

\paragraph{The Bayesian analysis.}

All that remains is to relate the loss incurred by our algorithm to
this drop. Indeed, we show precisely this for our Bayesian contextual
search algorithm: if the context vector is $\bx_t$, then our
algorithm's expected regret is at most a constant times the drop in
standard deviation (see the proof of
\Cref{thm:bayesian-contextual-search}{).} Using the above inequalities
and summing over all {time steps}, we get that the expected total regret
is at most {a constant times the expected total drop} in potential $\Psi$. Since the
potential is non-negative, this is at most {the same constant times the initial potential},
which is $\Tr
\sqrt{\Sigma}${.} 
{This completes} the Bayesian proof.

For contextual bilateral trade, our potential is the sum of two
{trace-sqrt potentials}---one for each of the unknown buyer and seller
vectors,
$\Psi_{\mathrm{tr}} = \Tr \sqrt{\Sigma_B}+\Tr \sqrt{\Sigma_S}$---and
we bound the {lost gain from trade} analogously.

\paragraph{Moving to the adversarial problem.}

Since the Bayesian problem has regret $O(\sqrt{d})$, it is tempting to
use the minimax theorem to infer the existence of an algorithm for the
adversarial case. 
We take a slightly different
route for contextual search. We go back to the proof of \Cref{lem:g-mono-concave}, where we
wrote $g_{\bx}(\Sigma)$ as the infimum of affine functions
in~(\ref{eq:affine}). A similar proof shows that
\begin{gather}\label{eq:variational-quadratic-Phi}
  2 \underbrace{\Tr\sqrt{\Sigma}}_{= \Psi(\Sigma)} \quad = \quad 
  \inf_{\bm\in\R^d,\,R\succ0}\mathbb E\big[\underbrace{\Tr R+(\bz-\bm)^\top
    R^{-1}(\bz-\bm)}_{=\; \Phi(\bm,R,\bz)}\bigr].
\end{gather}

The function $\Psi$ on the left was our Bayesian potential, whereas
$\Phi$ on the right is what we use as our potential for
the adversarial problem---the potentials for the ``dual''
Bayesian/adversarial problems are duals to each other!

We then show that our algorithm, which captures its uncertainty about
$\bz$ (which is now an adversarially chosen quantity) using the
posterior ``proxy mean'' $\bm_t$ and ``proxy covariance'' matrix
$R_t^2$, 
satisfies
\[ \text{expected cost of the algorithm at any time} \quad \leq \quad
  O(1)\cdot \text{expected change in $\Phi$}. \]
{For bilateral trade, we again sum the seller's and buyer's quadratic potentials. While in contextual search we can achieve the displayed bound through an explicit state update, in bilateral trade, a one-round minimax argument (\Cref{lem:trade-lottery}) instead guarantees a finite lottery of prices and successor states satisfying the bound for every hidden pair, up to arbitrarily small slack.}

\subsection{Related Work}

Contextual search was introduced as a multidimensional generalization of binary search \cite{PLS18}, providing a common framework {for} earlier contextual pricing problems at the interface of computer science and economics \cite{KleinbergL03,AminRS14,CLPL16,Javanmard17}.  Most existing approaches to contextual search are inspired by \textit{cutting-plane methods} in optimization, where the algorithm maintains a polytope of feasible targets that shrinks as the learner acquires information (in each round, a half-space of targets is incompatible with the feedback received). Since computing a candidate price based on the barycenter or Steiner point of the feasible polytope is insufficient \cite{LPS21}, \cite{PLS18} relied on tools from integral geometry. 
{The loss bound was improved to} $O(d\log d)$ in \cite{LPS21}, by posting a price corresponding to the barycenter of an augmented version of the feasible polytope. 

Our proxy covariance matrix approach is reminiscent of \textit{ellipsoid methods} in optimization, which maintain a quadratic approximation of the feasible region \cite{yudin1976informational,shor1977cut}. Ellipsoid methods have previously been applied to contextual search, e.g., in \cite{CLPL16,LPLV18,GGKMPS,GBCCP25}. A key novelty of our approach is that our proxy ellipsoid (or constant-factor dilatation of it) does not contain the feasible region. In fact, it expands in rounds where it receives ``exterior'' feedback! Of course, on average over several rounds, our proxy ellipsoid concentrates on the target parameters. 

{Related work on contextual search has} also explored other loss functions, such as the \textit{pricing loss}, which measures the revenue shortfall relative to knowing the buyer's valuation{, for which} \cite{LPS21} obtain $O(d\log\log T+d\log d)$ regret. {Another criterion counts predictions whose error exceeds a prescribed tolerance} $\epsilon${; for this criterion,} \cite{LPLV18} give an algorithm making at most $O(d\log(\nicefrac{d}{\epsilon}))$ errors. 

The geometric viewpoint also connects contextual search to inverse optimization and low-regret cutting-plane methods~\cite{Barmann17,GGKMPS,Besbes25,SakaueONS}. In online inverse linear optimization, the learner recommends an action from a feasible set and then observes the action that maximizes the unknown linear objective over that set. Under Euclidean unit-ball normalization, \cite{Sakaue26,gupta2026optimal} obtain a horizon-independent $O(\sqrt d)$ expected-regret bound for this problem. The distinction in feedback is important: the optimal action feedback is more informative than the one-bit contextual search feedback.

The problem of bilateral trade goes back to \cite{MS83}, which, interestingly, originally focused on a one-dimensional Bayesian setting somewhat related to ours, albeit with a single round $(T=1)$. Cesa-Bianchi et al.~\cite{CCCFL21,Cesa-BianchiCCF24} studied a sequential version of the problem where the algorithm competes with the best fixed price in hindsight, with valuations sampled from a fixed unknown distribution; here, the regret unavoidably depends on the number of rounds $T$, but the dependence can be made sublinear. The contextual (multi-dimensional) problem was introduced by \cite{GBCCP25}; their noiseless algorithm, which requires two-bit feedback, has $\tilde O(d^2\log T)$ regret. In the weaker one-bit feedback setting, the best time-independent result achieves $O(d\,6^d)$ regret \cite{CFGLPR26}. 

In variants of bilateral trade, the broker can post two distinct prices: $p$ for the buyer and $q$ for the seller. This extra degree of freedom can facilitate maximizing gain from trade. To restrict that advantage, it is usually assumed that the algorithm respects ``global budget balance'' \cite{BCCF24}, i.e., that it does not run persistent deficits,  or ``per-round budget balance'' \cite{Cesa-BianchiCCF24}, which imposes $q\leq p$ {in every round}. Allowing the broker to post distinct prices also enables other measures of performance for bilateral trade, such as broker profit maximization \cite{GBCCP25,DiGregorioDFS25}. Our algorithms do not use this flexibility.

\subsection{Notation}\label{sec:notation}

We use $\norm{\cdot}$ for the Euclidean norm, $B_2^d=\{\boldsymbol v\in\R^d:\norm{\boldsymbol v}\le1\}$ for the {Euclidean} ball, and $A\succeq B$ for the positive-semidefinite L\"owner order on symmetric matrices. The {positive-semidefinite square root} of a positive-semidefinite matrix is unique and denoted by $\Sigma^{\nicefrac12}$. We write $\Tr(\cdot)$ for trace, $\Cov(\cdot)$ for covariance, and $\sd(\cdot)$ for standard deviation. We set $\operatorname{sign}(0)=1${, and write} $a_+=\max\{a,0\}$ {for the positive part}.

We usually use bold symbols to denote vectors in $\Rbb^d$. Specifically, we use $\bmu$ for a distributional mean and $\bm$ for the algorithmic mean proxy, with scalar projections $\mu=\bx^\top\bmu$ and $m=\bx^\top\bm$ along a context $\bx$. A covariance matrix is denoted by $\Sigma$, with directional standard deviation $\sigma=\sqrt{\bx^\top\Sigma\bx}$. 

The observed history at time $t$, denoted by $\mathcal F_{t-1}$, contains the contexts, posted prices, feedback, and the learner's randomization from previous rounds. We write $\Et=\E{\,\cdot\mid\mathcal F_{t-1},\bx_t}$ for expectation conditional on this history and the current context, before the learner's fresh randomization. For both contextual search and bilateral trade, contexts may adapt to this history but are chosen before the learner's fresh randomization. In the Bayesian setting, each context is conditionally independent of the hidden vector or pair given $\mathcal F_{t-1}$, so observing it does not change the posterior.

Additional notation is summarized in the table in Appendix~\ref{ap:nota}.

\section{Contextual Search}\label{sec:search}

In contextual search, the learner is looking for an unknown target $\bz\in\R^d$. At round $t$, the learner receives a context $\bx_t$ with $\norm{\bx_t}\le1$, posts $P_t\in\R$, suffers loss 
\[
 \ell_t=|\bx_t^\top\bz-P_t|, \qquad\text{and receives feedback}\qquad Y_t=\operatorname{sign}(\bx_t^\top\bz-P_t),
\]
without observing either the target value or the loss.  We seek bounds on the total loss $\sum_{t=1}^T\E{\ell_t}$ that are independent of $T$.
In this section, we first present a Bayesian algorithm whose expected total loss is at most $24\,\Tr\sqrt{\Sigma}$ for any known prior on $\bz$ with finite second moment and covariance matrix $\Sigma$ (Theorem~\ref{thm:bayesian-contextual-search}). We then give an algorithm that needs no prior and has expected total loss at most $12\sqrt d$ when the hidden vector $\bz$ lies in the Euclidean unit ball (Theorem~\ref{thm:main}). Both guarantees hold for every horizon $T$. For completeness, we recall the classical $\Omega(\sqrt d)$ lower bound for contextual search in Appendix~\ref{app:lower-bound}.

\subsection{Bayesian Contextual Search}\label{sec:bayes}\label{sec:gap}

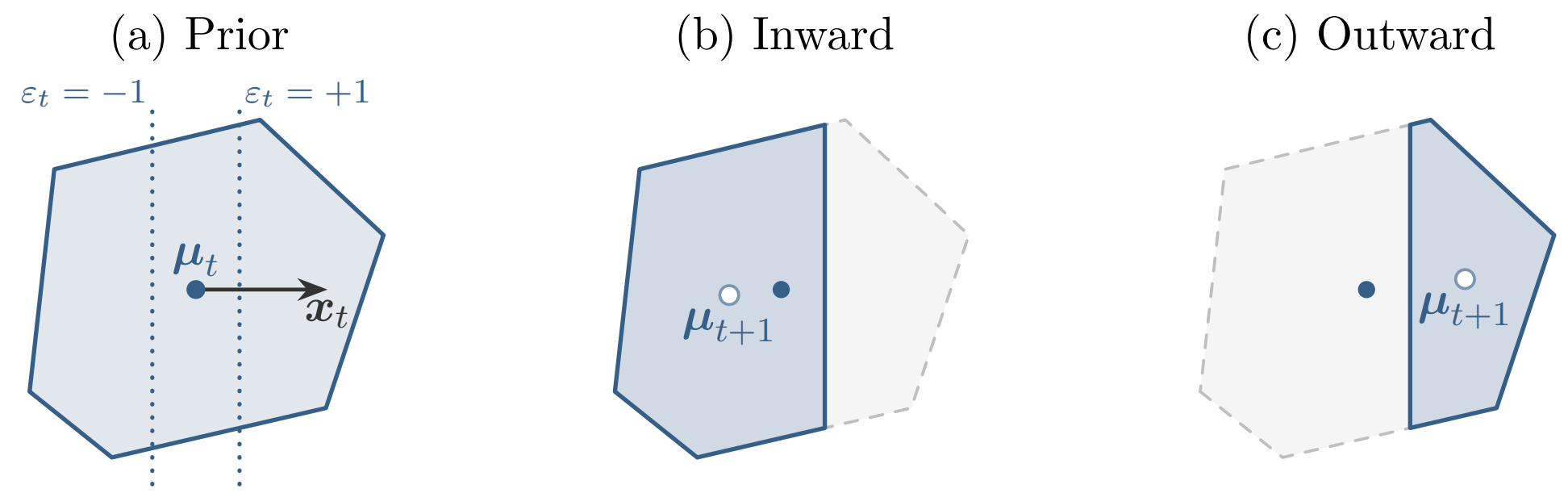
\begin{figure}[t]
\centering
\begin{tikzpicture}[
  x=1.35cm, y=1.35cm,
  line cap=round, line join=round,
  >=Stealth
]
\definecolor{regionblue}{RGB}{54,95,136}
\definecolor{cutorange}{RGB}{180,85,35}

\def\priorpolygon{(-1,-.6)--(-.5,-1)--(.8,-.7)--(1.15,.35)--(.4,1.05)--(-.85,.75)--cycle}
\def\posteriorpolygon{(-1,-.6)--(-.5,-1)--(.2751461509,-.8211201190)--(.2751461509,1.0200350762)--(-.85,.75)--cycle}
\def\outerposteriorpolygon{(.2751461509,-.8211201190)--(.8,-.7)--(1.15,.35)--(.4,1.05)--(.2751461509,1.0200350762)--cycle}
\def\cuttop{(.2751461509,1.1)}
\def\cutbottom{(.2751461509,-1.2)}
\def\cutminustop{(-.2547227065,1.1)}
\def\cutminusbottom{(-.2547227065,-1.2)}

\begin{scope}
  \node at (.03,1.55) {(a) Prior};
  \path[fill=regionblue!15,draw=regionblue,line width=1pt] \priorpolygon;

  \draw[regionblue,dash pattern=on 0pt off 3pt,line width=1pt] \cutminustop--\cutminusbottom;
  \draw[regionblue,dash pattern=on 0pt off 3pt,line width=1pt] \cuttop--\cutbottom;
  \node[anchor=south east,font=\scriptsize,inner sep=1pt,text=regionblue]
    at \cutminustop {$\varepsilon_t=-1$};
  \node[anchor=south west,font=\scriptsize,inner sep=1pt,text=regionblue]
    at \cuttop {$\varepsilon_t=+1$};

  \draw[->,black!80,line width=.9pt]
    (.0102117222,.0193777434)--(.8102117222,.0193777434)
    node[below,inner sep=2pt] {$\bx_t$};
  \fill[regionblue] (.0102117222,.0193777434) circle[radius=2.2pt];
  \node[anchor=south,inner sep=3pt,text=regionblue]
    at (.0102117222,.0193777434) {$\bmu_t$};
\end{scope}

\foreach \panelshift/\answer/\panelletter in {4.8/-1/b,9.6/1/c}{
  \begin{scope}[xshift=\panelshift cm]
    \ifnum\answer=-1
  \node at (.03,1.55) {(\panelletter) Inward};
\else
  \node at (.03,1.55) {(\panelletter) Outward};
\fi
    \path[fill=black!4,draw=black!25,dashed,line width=.7pt] \priorpolygon;
    \ifnum\answer=-1
      \path[fill=regionblue!23,draw=regionblue,line width=1pt] \posteriorpolygon;
    \else
      \path[fill=regionblue!23,draw=regionblue,line width=1pt] \outerposteriorpolygon;
    \fi

    \fill[regionblue]
      (.0102117222,.0193777434) circle[radius=2pt];
    \ifnum\answer=-1
      \draw[fill=white,draw=regionblue!65,line width=.7pt] (-.3055486711,-.0141238531) circle[radius=2.2pt];
      \node[anchor=north,inner sep=3pt,text=regionblue]
        at (-.3055486711,-.0141238531) {$\bmu_{t+1}$};
    \else
      \draw[fill=white,draw=regionblue!65,line width=.7pt] (.6087370913,.0828801833) circle[radius=2.2pt];
      \node[anchor=north,inner sep=3pt,text=regionblue]
        at (.6087370913,.0828801833) {$\bmu_{t+1}$};
    \fi
  \end{scope}
}
\end{tikzpicture}
\caption{{(a) Bayesian prior; (b) posterior for $\varepsilon_t=1$ with $Y_t=-1$; (c)
posterior for $\varepsilon_t=1$ with $Y_t=1$.}}
\label{fig:bayesian-polygon-update}
\end{figure}

In the Bayesian setting, the target $\bz\in\R^d$ has a known prior with finite second moment. That distribution is updated with the information {acquired} by the learner in previous rounds. Just before round $t$, we denote by
\[
 \bmu_t=\E{\bz\mid\mathcal F_{t-1}},
 \qquad \Sigma_t=\Cov(\bz\mid\mathcal F_{t-1})
\]
the posterior mean and covariance of the target. For the current context $\bx_t$, we denote the scalar target by $Z=\bx_t^\top\bz$, its conditional mean by $\mu_t=\bx_t^\top\bmu_t$ and its standard deviation by $\sigma_t=\agsqrt{\bx_t^\top\Sigma_t\bx_t}$. Algorithm~\ref{alg:bayesian} queries one of two prices centered at $\mu_t$, as illustrated in Figure~\ref{fig:bayesian-polygon-update}.

\begin{algorithm}[H]
\caption{Bayesian Contextual Search}\label{alg:bayesian}
\vspace{3pt}
Given a known prior on $\bz$ with finite second moment, repeat the following at each round $t=1,2,\ldots$:
\begin{enumerate}
 \item Compute the posterior moments $\bmu_t,\Sigma_t$ of $\bz$ and their projections on the context vector $\bx_t$,
 \[
   \mu_t=\bx_t^\top\bmu_t,\qquad
   \sigma_t=\agsqrt{\bx_t^\top\Sigma_t\bx_t}.
 \]
 \item Draw a fresh independent uniform sign $\varepsilon_t\in\{-1,1\}$ and post the price
 \[
   P_t=\mu_t+\frac{\sigma_t}{2}\varepsilon_t.
 \]
 \item Observe $Y_t=\operatorname{sign}(\bx_t^\top\bz-P_t)$ and compute the posterior distribution on $\bz$.
\end{enumerate}
\end{algorithm}

\begin{theorem}[Bayesian Contextual Search]\label{thm:bayesian-contextual-search}
For any known prior on $\bz\in\R^d$ with finite second moment and covariance $\Sigma_1$, Algorithm~\ref{alg:bayesian} satisfies, for every horizon $T\ge1$,
\begin{equation}\label{eq:bayesian-guarantee}
 \sum_{t=1}^T\E{\ell_t}\le24\,\Tr\sqrt{\Sigma_1}.
\end{equation}
{The expectation is over the prior and the randomization of the learner and the contexts.}
\end{theorem}

\begin{proof}[Proof of Theorem~\ref{thm:bayesian-contextual-search}]
  {We use the following one-dimensional contraction bound for the standard deviation, stated in \Cref{lem:deviation-contraction}.} The lemma,
  whose proof is elementary, implies that a single bit of information
  suffices to shrink the posterior standard deviation of a scalar
  distribution by a constant factor. The proof of this lemma is given
  in Appendix~\ref{app:contraction}.

\begin{restatable}[One-bit deviation contraction]{lemma}{deviationcontraction}\label{lem:deviation-contraction}
Let $Z$ be a square-integrable real random variable, with $\mu=\E{Z}$ and $\sigma=\sd(Z)$. Let $\varepsilon$ be uniform on $\{-1,1\}$ and independent of $Z$. {Define} 
\[
 P=\mu+\frac\sigma2\varepsilon,
 \qquad Y=\operatorname{sign}(Z-P).
\]
{Writing the posterior standard deviation as} $\sigma'=\sd(Z\mid P,Y)$, we have
\(
\E{\sigma'}\le\frac{23}{24}\sigma.
\)
\end{restatable}
The intuition is the following: because the threshold is placed at $\pm \frac{1}{2}\sigma$ away from the mean, the middle region ${[\mu-\frac{1}{2}\sigma, \mu+\frac{1}{2}\sigma]}$ can hold at most $\frac{1}{4}$ of the variance $\sigma^2$; at least one of the two outer tails must hold a large fraction of the variance. Conditioning on whether $Z$ falls in that tail creates {a win-win situation:}
\begin{itemize}
    \item If the tail has large probability, then every point in the tail is at least $\frac{1}{2}\sigma$ away from the mean, so learning whether $Z$ is in the tail causes a large shift in the posterior mean, which reduces the variance (by the law of total variance).
\item If the tail has small probability, then isolating the tail removes a large chunk of variance from the ``main body'' of the distribution (shrinking the main body's standard deviation), while the tail itself {occurs} so rarely that its own conditional standard deviation—which grows only like $\nicefrac{1}{\sqrt{p}}$ while its probability shrinks like $p$—contributes almost nothing ($\propto \sqrt{p}$) to the expectation.
\end{itemize}

We now move on to the core argument. First, we bound the loss in a round of \Cref{alg:bayesian} by $\sigma_t$, the uncertainty along the context vector{. If $\sigma_t=0$, then $P_t=Z=\mu_t$ almost surely, so the loss is zero. For $\sigma_t>0$, we have}
\[
 \Esub{t}{\ell_t}
 =\Esub{t}{|Z-P_t|}=\Esub{t}{\max\left\{|Z-\mu_t|,\frac{\sigma_t}{2}\right\}}
 \le\frac{\sigma_t}{2}+\frac{\Esub{t}{(Z-\mu_t)^2}}{2\sigma_t}
 =\sigma_t,
\]
where we used $\max\{|u|,\nicefrac{\sigma_t}2\}\le\nicefrac{\sigma_t}2+\nicefrac{u^2}{2\sigma_t}$. 

Next, we pay for each round's loss with a decrease in the posterior potential
 $\Psi_t=\Tr\sqrt{\Sigma_t}$, where $\Sigma_t=\Cov(\bz\mid\mathcal F_{t-1})$.
Let
$\sigma_t=\agsqrt{\bx_t^\top\Sigma_t\bx_t}$ and
$\sigma_t'=\agsqrt{\bx_t^\top\Sigma_{t+1}\bx_t}$ be the directional standard deviations
of the target projection before and after the observation; we apply \Cref{lem:deviation-contraction} to the random variable $\bx_t^T\bz$ and  \Cref{lem:directional-sd-decrease} to the potential $\Psi_t$ to obtain {the potential inequality}
\[
 \Esub{t}{\ell_t}\le\sigma_t
 \overset{\text{Lemma~\ref{lem:deviation-contraction}}}{\le}
 24\bigl(\sigma_t-\Esub{t}{\sigma_t'}
 \bigr)
 \overset{\text{Lemma~\ref{lem:directional-sd-decrease}}}{\le}
 24\bigl(\Psi_t-\Esub{t}{\Psi_{t+1}}\bigr).
\]
{Telescoping this inequality yields the theorem:}
\[
 \sum_{t=1}^T\E{\ell_t}
 \le
 24\bigl(\Psi_1-\E{\Psi_{T+1}}\bigr)
 \le24\Psi_1.\qedhere
\]
\end{proof}

\subsection{Adversarial Contextual Search}\label{sec:adversarial}\label{sec:algorithm}

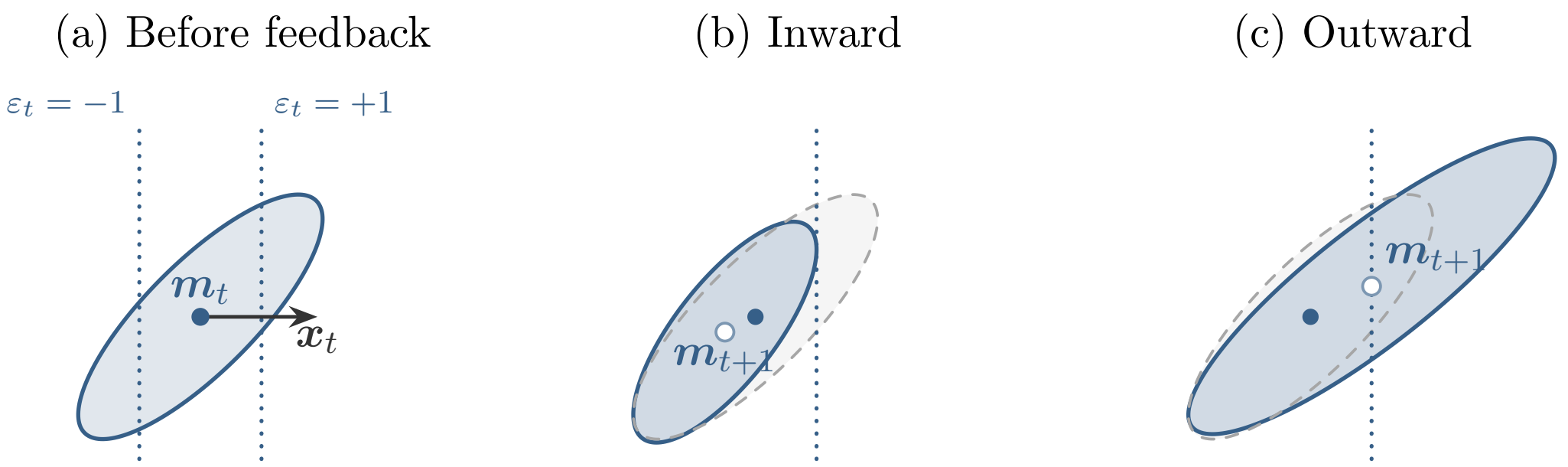
\begin{figure}[tbp]
\centering
\begin{tikzpicture}[
  x=1.35cm, y=1.35cm,
  line cap=round, line join=round,
  >=Stealth
]
\definecolor{regionblue}{RGB}{54,95,136}
\definecolor{cutorange}{RGB}{180,85,35}

\def\wcRxx{.7}
\def\wcRxy{.35}
\def\wcRyy{.7}
\def\wcXx{1}
\def\wcXy{0}
\pgfmathsetmacro{\wcUx}{\wcRxx*\wcXx+\wcRxy*\wcXy}
\pgfmathsetmacro{\wcUy}{\wcRxy*\wcXx+\wcRyy*\wcXy}
\pgfmathsetmacro{\wcA}{\wcXx*\wcUx+\wcXy*\wcUy}
\pgfmathsetmacro{\wcRadius}{sqrt(\wcUx*\wcUx+\wcUy*\wcUy)}
\pgfmathsetmacro{\wcPrice}{\wcRadius/2}

\foreach \panelshift/\panelstate/\panelletter/\paneltitle in
  {0/0/a/Before feedback,4.8/-1/b/Inward,9.6/1/c/Outward}{
  \begin{scope}[xshift=\panelshift cm]
    \ifnum\panelstate=0
      \node at (.27,1.8)
        {(\panelletter) \paneltitle};
    \else
      \pgfmathsetmacro{\wcFactor}{\panelstate/(\wcA*(3-\panelstate))}
      \pgfmathsetmacro{\wcMx}{\wcFactor*\wcRadius*\wcUx}
      \pgfmathsetmacro{\wcMy}{\wcFactor*\wcRadius*\wcUy}
      \pgfmathsetmacro{\wcNewRxx}{\wcRxx+\wcFactor*\wcUx*\wcUx}
      \pgfmathsetmacro{\wcNewRxy}{\wcRxy+\wcFactor*\wcUx*\wcUy}
      \pgfmathsetmacro{\wcNewRyy}{\wcRyy+\wcFactor*\wcUy*\wcUy}
      \node at (.27,1.8)
        {(\panelletter) \paneltitle};
    \fi
    \ifnum\panelstate=0
      \path[fill=regionblue!15,draw=regionblue,line width=1pt]
        plot[domain=0:360,samples=181,variable=\angle]
          ({\wcRxx*cos(\angle)+\wcRxy*sin(\angle)},
           {\wcRxy*cos(\angle)+\wcRyy*sin(\angle)}) -- cycle;
    \else
      \path[fill=black!4]
        plot[domain=0:360,samples=181,variable=\angle]
          ({\wcRxx*cos(\angle)+\wcRxy*sin(\angle)},
           {\wcRxy*cos(\angle)+\wcRyy*sin(\angle)}) -- cycle;
      \path[fill=regionblue!23,draw=regionblue,line width=1pt]
        plot[domain=0:360,samples=181,variable=\angle]
          ({\wcMx+\wcNewRxx*cos(\angle)+\wcNewRxy*sin(\angle)},
           {\wcMy+\wcNewRxy*cos(\angle)+\wcNewRyy*sin(\angle)}) -- cycle;
      \draw[black!35,dashed,line width=.7pt]
        plot[domain=0:360,samples=181,variable=\angle]
          ({\wcRxx*cos(\angle)+\wcRxy*sin(\angle)},
           {\wcRxy*cos(\angle)+\wcRyy*sin(\angle)}) -- cycle;
    \fi

    \ifnum\panelstate=0
      \foreach \coin in {-1,1}{
        \draw[regionblue,dash pattern=on 0pt off 3pt,line width=1pt]
          ({\coin*\wcPrice},-.91)--({\coin*\wcPrice},1.20);
      }
      \node[anchor=south east,inner sep=3pt,text=regionblue, font = \scriptsize]
        at (-\wcPrice,1.20) {$\varepsilon_t=-1$};
      \node[anchor=south west,inner sep=3pt,text=regionblue, font=\scriptsize]
        at (\wcPrice,1.20) {$\varepsilon_t=+1$};
    \else
      \draw[regionblue,dash pattern=on 0pt off 3pt,line width=1pt]
        (\wcPrice,-.91)--(\wcPrice,1.20);
    \fi
    \ifnum\panelstate=0
    \draw[->,black!80,line width=.9pt]
      (0,0)--({.75*\wcXx},{.75*\wcXy})
      node[below,inner sep=2pt]{$\bx_t$};
    \fi 
    
    \ifnum\panelstate=0
      \fill[regionblue] (0,0) circle[radius=2.2pt];
      \node[anchor=south,inner sep=3pt,text=regionblue]
        at (0,0) {$\bm_t$};
    \else
      \fill[regionblue]
        (0,0) circle[radius=2pt];
       \draw[fill=white,draw=regionblue!65,line width=.7pt] (\wcMx,\wcMy) circle[radius=2.2pt];
      \ifnum\panelstate=-1
        \node[anchor=north,inner sep=3pt,text=regionblue]
          at (\wcMx,\wcMy) {$\bm_{t+1}$};
      \else
        \node[anchor=south west,inner sep=3pt,text=regionblue]
          at (\wcMx,\wcMy) {$\bm_{t+1}$};
      \fi
    \fi
  \end{scope}
}
\end{tikzpicture}
\caption{{Proxy ellipsoids: (a) before feedback; (b) after $\varepsilon_t=1$ with $Y_t = -1$; (c) after $\varepsilon_t=1$
with $Y_t = 1$.}}
\label{fig:worst-case-scale-update}
\end{figure}

We now consider a fixed unknown target $\bz$ in the unit Euclidean ball. Our worst-case algorithm, \Cref{alg:adversarial}, maintains two parameters: a vector $\bm_t\in\R^d$, the mean proxy, and a symmetric positive-definite scale matrix $R_t\in\R^{d\times d}$, the square-root covariance proxy.
This algorithm mirrors the Bayesian algorithm, \Cref{alg:bayesian}, in that $\bm_t$ plays the role of $\bmu_t$ and $R_t$ plays the role of $\Sigma_t^{\nicefrac 12}$ (their one-dimensional projections along the context axis are denoted by $m_t$ and $r_t$). The key nuance is that, without Bayesian posterior updates, \Cref{alg:adversarial} specifies explicit updates for both proxies $\bm_t,R_t$ that depend only on the posted price and {the feedback received, as specified in} \eqref{eq:update}. Figure~\ref{fig:worst-case-scale-update} illustrates one round of the algorithm.

\begin{algorithm}[H]
\caption{Adversarial Contextual Search}\label{alg:adversarial}
\vspace{3pt}
Initialize
\begin{equation}\label{eq:init}
 \bm_1=0,\qquad R_1=\frac1{\sqrt d}I_d.
\end{equation}
At each round $t=1,2,\ldots$:
\begin{enumerate}
 \item Receive $\bx_t$, compute
 \(
 m_t=\bx_t^\top\bm_t,\,
 \boldsymbol u_t=R_t\bx_t,\,
 a_t=\bx_t^\top\boldsymbol u_t,\) and \(
 r_t=\norm{\boldsymbol u_t}.
 \)
 \item Draw a fresh independent uniform {sign }$\varepsilon_t\in\{-1,1\}$ and post
 \begin{equation}\label{eq:price}
 P_t=m_t+\frac{r_t}{2}\varepsilon_t.
 \end{equation}
 \item Observe $Y_t$ and update
 \begin{equation}\label{eq:update}
 \bm_{t+1}=\bm_t+\frac{Y_t r_t}{a_t(3-\varepsilon_t Y_t)}\boldsymbol u_t,
 \qquad
 R_{t+1}=R_t+\frac{\varepsilon_t Y_t}{a_t(3-\varepsilon_t Y_t)}\boldsymbol u_t\boldsymbol u_t^\top.
 \end{equation}
\end{enumerate}
\end{algorithm}

Before providing our regret guarantee, we first verify that the algorithm is well-defined, {meaning that} the updates in \eqref{eq:update} preserve the positive definiteness of the scale matrix $R_t\succ 0$. This observation is obvious in the case where $\varepsilon_t Y_t = 1$, because {adding a positive semidefinite matrix preserves positive definiteness}; so it suffices to verify that {positive definiteness} is preserved whenever $\varepsilon_t Y_t = - 1$.

\begin{remark}[$R_t\succ 0$ throughout execution]\label{rem:pd}
{Assume $R\succ0$ and $\bx\ne0$, and write $\boldsymbol u=R\bx$ and $a=\bx^\top R\bx$.} Cauchy--Schwarz in the inner product induced by $R$ gives
\[
 (\boldsymbol v^\top\boldsymbol u)^2
 \le(\boldsymbol v^\top R\boldsymbol v)(\bx^\top R\bx)
 \quad\text{for every }\boldsymbol v\in\R^d.
\]
Hence $\boldsymbol u\boldsymbol u^\top/a\preceq R$, and the update rule in \eqref{eq:update} satisfies
\[
 R-\frac{\boldsymbol u\boldsymbol u^\top}{4a}\succeq\frac34R\succ0.
\]
{This proves the claim by induction over the rounds.}
\end{remark}

\begin{theorem}[Adversarial Contextual Search]\label{thm:main}
For every fixed $\bz$ with $\norm{\bz}\le1$, every horizon $T\ge1$, and every admissible adaptive context strategy with $\norm{\bx_t}\le1$, Algorithm~\ref{alg:adversarial} satisfies
\begin{equation}\label{eq:adversarial-guarantee}
 \sum_{t=1}^T\E{\ell_t}\le12\sqrt d.
\end{equation}
The algorithm uses $O(d^2)$ arithmetic operations and $O(d^2)$ real-valued storage per round.
\end{theorem}

\begin{proof}[Proof of Theorem~\ref{thm:main}]
For a fixed hidden target, we consider the non-negative potential defined by
\begin{equation}\label{eq:potential}
 \Phi_t=\Phi(\bm_t,R_t,\bz)
 =\Tr R_t+(\bz-\bm_t)^\top R_t^{-1}(\bz-\bm_t).
\end{equation}
At initialization, the potential {satisfies}
\(
\Phi_1=\sqrt d+\sqrt d\,\norm{\bz}^2\le2\sqrt d.
\)
We use the following potential-decrease bound, which is shown in
Appendix~\ref{app:drift}. 
\begin{restatable}[One-round potential decrease]{lemma}{potentialdecrease}\label{lem:drift}
On every round of Algorithm~\ref{alg:adversarial},
\begin{equation}\label{eq:drift}
 \Esub{t}{\ell_t}\le6\bigl(\Phi_t-\Esub{t}{\Phi_{t+1}}\bigr).
\end{equation}
\end{restatable}

{As in the Bayesian case, the proof reduces to one dimension:} we define
$\varphi(m,r,z)=r+\frac{(m-z)^2}{r}$, which is just a scalar version
of the potential $\Phi(\cdot)$. Going from $R_t$ to $R_{t+1}$ is a
rank $1$ perturbation, {so the Sherman--Morrison formula expresses the expected potential drop}
$\Phi_t-\Esub{t}{\Phi_{t+1}}${\ as a multiple of}
$\varphi(m_t,r_t,z) - \Esub{t}{\varphi(m_{t+1},r_{t+1},z)}$, where $m_t$ and
$r_t$ are the projected mean and deviation proxies; 
the proof now follows from elementary calculations. 

Summing the {potential inequality} \eqref{eq:drift} over rounds yields
\(
 \sum_{t=1}^T\E{\ell_t}+6{\E{\Phi_{T+1}}}
 \le6\Phi_1\le12\sqrt d,
\)
which proves the theorem. 

In the algorithm, computing $R_t\bx_t$ and performing the rank-one update each take $O(d^2)$ arithmetic operations, whereas storing $R_t$ and $\bm_t$ requires $O(d^2)$ real numbers. Importantly, the algorithm never evaluates $\Phi_t$ or computes $R_t^{-1}$, which only serve in the analysis. 
\end{proof}

\section{Contextual Bilateral Trade}\label{sec:trade}\label{sec:trade-model}

In bilateral trade, a broker intermediates between a seller and a buyer who are represented by two vectors $\boldsymbol s,\boldsymbol b\in\R^d$. {At round $t$, the broker receives a context vector $\bx_t$ with $\norm{\bx_t}\le1$, proposes a price $p_t$, and observes whether trade occurs:}
\[
 Y_t=\mathbf1\{\bx_t^\top\boldsymbol s\le p_t\le\bx_t^\top\boldsymbol b\}.
\]
In particular, a failed trade does not identify the agent who rejected the price. The lost ``gain from trade'' in round $t$ is defined as 
\(
 \ell_t=(\bx_t^\top(\boldsymbol b-\boldsymbol s))_+(1-Y_t).
\)

In \Cref{sec:trade-bayes} we present a Bayesian algorithm whose expected total loss is at most {$2^{11}\bigl(\Tr\sqrt{\Sigma_S}+\Tr\sqrt{\Sigma_B}\bigr)$}, for any known prior on the targets with marginal covariance matrices $\Sigma_S$ and $\Sigma_B$ (\Cref{thm:trade-bayes}). In \Cref{sec:trade-direct}, we
then give an algorithm that needs no prior and has expected total loss at most ${2^{12}\sqrt d+1}$ when both targets lie in the unit Euclidean {ball, so that} $\boldsymbol w=(\boldsymbol s,\boldsymbol b)\in K=\mathbb B_2^d\times\mathbb B_2^d$ (\Cref{thm:trade-direct}). All our guarantees hold for every horizon $T$. For completeness, we recall the classical $\Omega(\sqrt d)$ lower bound for bilateral trade in Appendix~\ref{app:lower-bound}.

\subsection{Bayesian Contextual Bilateral Trade}\label{sec:trade-bayes}

In the Bayesian setting, write $W=(\bs,\bb)$ for the random pair of valuations. Its components may be dependent, and we require only finite second moments. For a distribution $\pi$ of $W$, define 
\begin{equation}\label{eq:trade-bayesian-potential}
 \Psi_{\mathrm{tr}}(\pi)
 =\Psi(\Cov(\bs))+\Psi(\Cov(\bb))
 =\Tr\sqrt{\Cov(\bs)}+\Tr\sqrt{\Cov(\bb)}.
\end{equation}
This is the sum of the two search potentials, not the trace square root of the joint covariance. If $\pi$ is supported on $K$, then $\Psi_{\mathrm{tr}}(\pi)\le2\sqrt d$, because each marginal covariance has trace at most one. For support in $[-1,1]^d\times[-1,1]^d$, the corresponding bound is $2d$.

Given a context $\bx$, the one-dimensional problem along $\bx$ now involves a pair $(S,B)$ rather than a single target. For a common price $p\in\R$, {let} $Y(p)=\mathbf1\{S\le p\le B\}$ {denote }the trade bit, so that the loss incurred by posting price $p$ is
\begin{align}
 L(p)=\E{(B-S)_+(1-Y(p))}.\label{eq:trade-scalar-loss}
\end{align}
{To quantify the information obtained from the trade bit}, for each $Z \in \{S,B\}$, we define the standard deviation drop $D_Z(p)=\sigma_Z-\E{\sd(Z\mid Y(p))}$, so that the overall drop is
\begin{align}
    D(p)&=D_S(p)+D_B(p).\label{eq:trade-scalar-information}
\end{align}
Each $D_Z(p)$ is nonnegative by {the law of total variance and the concavity} of the square root. We now introduce the key lemma that shows the existence of a specific price $p$ relating these two terms. We provide a proof sketch below, while the full proof is deferred to Appendix~\ref{sec:trade-toi}.

\begin{restatable}[Trade or information]{lemma}{tradeorinformation}\label{lem:toi}
For every pair $(S,B)$ of square-integrable real random variables, there is a common price $p$ such that
\begin{equation}\label{eq:toi}
 L(p)\le2^{11}D(p).
\end{equation}
The price can be chosen as a function of the joint distribution.
\end{restatable}

\begin{figure}[t!]
\centering
\begin{tikzpicture}[x=0.98cm,y=0.98cm,>=Stealth,
 declare function={
   sig(\x)=1/(1+exp(-max(min(\x,8),-8)));
   prof(\x)=0.85*sig((\x-2)/0.22)*sig((5-\x)/0.22)
            +0.035*exp(-((\x+0.5)/1.3)^2)
            +0.035*exp(-((\x-6.3)/1.0)^2);
 }]
 \definecolor{regionblue}{RGB}{54,95,136}
 \definecolor{cutorange}{RGB}{180,85,35}
 \def\H{2.8}
 \draw[->] (-4,0)--(9.8,0) node[right] {$p$};
 \draw[->] (-4,0)--(-4,3.4) node[above] {$\alpha(p)$};
 \draw[dashed,gray] (-4,\H)--(9.6,\H) node[right,black] {$1$};
 \fill[regionblue!15] plot[domain=-4:9.6,samples=240,smooth]
   (\x,{\H*prof(\x)})--(9.6,0)--(-4,0)--cycle;
 \draw[regionblue,thick] plot[domain=-4:9.6,samples=240,smooth]
   (\x,{\H*prof(\x)});
 \node[regionblue] at (3.5,1.3) {$G=\displaystyle\int\alpha$};
 \draw[cutorange,thick,densely dashed]
   plot[domain=-4:-1.4,samples=60] (\x,{\H*0.064*(1.4/\x)^2})
   --(-1.4,{\H*0.4})--(1.4,{\H*0.4})--(1.4,\H)
   --(5.6,\H)--(5.6,{\H*0.622222})--(7.4,{\H*0.622222})
   --(7.4,{\H*0.099556})
   plot[domain=7.4:9.6,samples=60]
     (\x,{\H*0.099556*(0.9/(\x-6.5))^2});
 \node[cutorange,above] at (-2.8,0.45)
   {$\dfrac{2\sigma_S\overline D}{(\mu_S-p)^2}$};
 \node[cutorange,above] at (0,{\H*0.4})
   {$\dfrac{225\overline D}{2\sigma_S}$};
 \node[cutorange,above] at (6.5,{\H*0.622222})
   {$\dfrac{225\overline D}{2\sigma_B}$};
 \node[cutorange,above] at (8.7,0.45)
   {$\dfrac{2\sigma_B\overline D}{(p-\mu_B)^2}$};
 \node[cutorange,above] at (3.5,\H) {$\alpha\le1$};
 \foreach \x in {-1.4,0,1.4,5.6,6.5,7.4}
   \draw (\x,0.07)--(\x,-0.07);
 \node[below] at (0,-0.05) {$\mu_S$};
 \node[below] at (6.5,-0.05) {$\mu_B$};
 \draw[<->] (-1.4,-0.65)--(1.4,-0.65)
   node[midway,below] {$|p-\mu_S|\le\nicefrac{\sigma_S}3$};
 \draw[<->] (5.6,-0.65)--(7.4,-0.65)
   node[midway,below] {$|p-\mu_B|\le\nicefrac{\sigma_B}3$};
\end{tikzpicture}
\caption{A schematic trade profile $\alpha$ (solid blue) and its piecewise upper bound $H$ (dashed orange) in the overlapping case. The area under $\alpha$ is $G$. The tails and central intervals contribute at most $162\overline D$ under $H$; the remaining plateau has area $(m-\nicefrac{\sigma}3)_+$. Claim~\ref{cl:trade-profile} establishes the bound.}
\label{fig:trade-profile}
\end{figure}
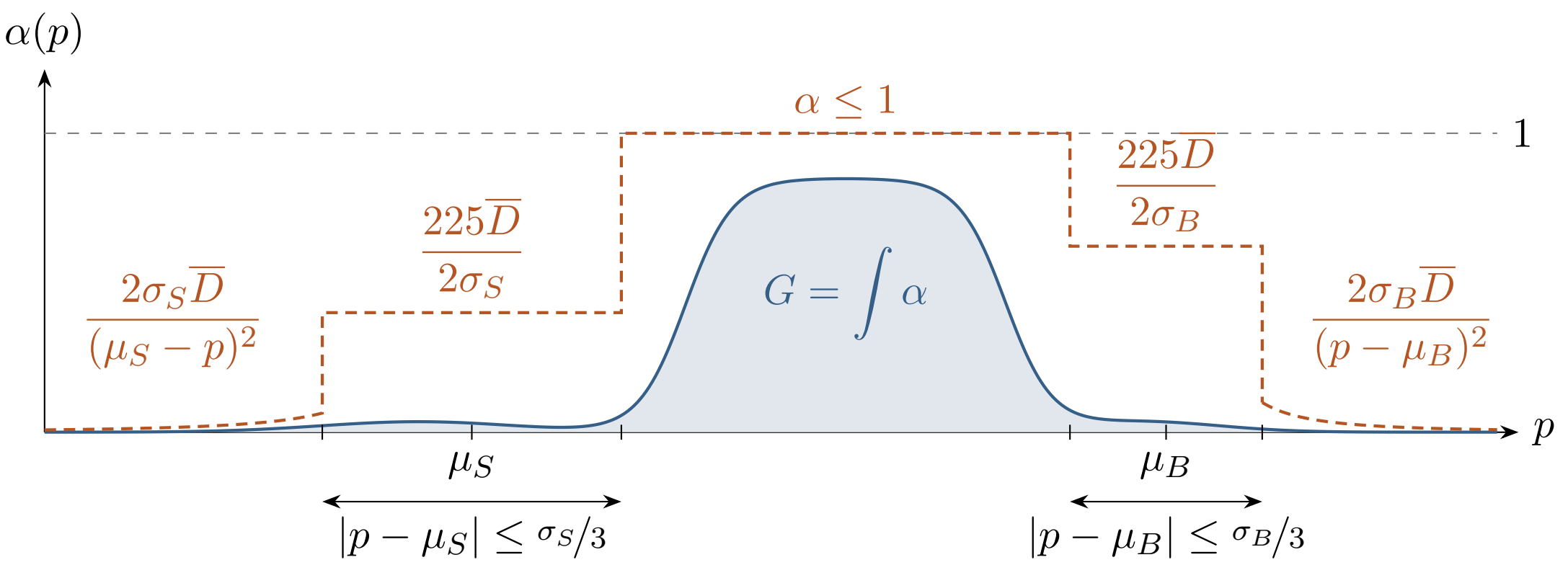

\begin{proof}[Proof sketch]
Let $\alpha(p)=\mathbb P(S\le p\le B)$ be the trade profile. For fixed valuations, $(B-S)_+$ is the length of the interval of prices permitting trade. Averaging this length and interchanging expectation and integration gives
\[
 G=\E{(B-S)_+}=\int_{\R}\alpha(p)\,dp.
\]
Write $\mu_Z={\E{Z}}$, $\sigma_Z=\sd(Z)$ for $Z \in \{S,B\}$, $\sigma=\sigma_S+\sigma_B$, and $m=\mu_B-\mu_S$. A deterministic valuation (i.e., $\sigma_Z = 0$) is handled by posting that value. Otherwise, the analysis has two cases.

\emph{Overlapping case ($m<3\sigma$).} In this case, the seller's and buyer's means are close to each other. To deal with this case, we put $\overline D=\sup_pD(p)$ and, since $L(p)\le G$, we bound the entire area under $\alpha$ by constructing the envelope $H$ in Figure~\ref{fig:trade-profile}. The crucial ingredient that we use to construct $H$ is the following information identity, which is proved in the appendix and holds for any $Z \in \{S,B\}$:
\[
	\E{\left(\E{Z\mid Y(p)}-\mu_Z\right)^2 +\bigl(\sd(Z\mid Y(p))-\sigma_Z\bigr)^2} =2\sigma_ZD_Z(p). \tag{Information Identity}
\]
The left-hand side averages the squared change in the mean and standard deviation after observing rejection or trade. Thus a feedback outcome that substantially changes either quantity must provide information, weighted by its probability. In the tails of $\alpha(\cdot)$, a successful trade forces a large shift of the conditional mean, yielding inverse-square bounds. For instance, 
{if} $p<\mu_S$, trading implies ${\E{S\mid Y(p)=1}}\le p$, thus providing {a} uniform upper bound on $\alpha(p)$ in terms of $D_S(p)$ (a similar derivation holds for the buyer's tail):
\begin{equation}
    	\alpha(p)(\mu_S-p)^2\le2\sigma_SD_S(p) \implies \alpha(p)\le \frac{2\sigma_S D(p)}{(\mu_S-p)^2}. \tag{Tail Bound}
\end{equation}

    For $|p-\mu_S|\le\nicefrac{\sigma_S}3$, consider the {values of sellers who trade} at $p$: they either lie within $\nicefrac{\sigma_S}3$ {of their mean or fall below} $\mu_S-\nicefrac{\sigma_S}3$ ({and would also permit trade at price} $\mu_S-\nicefrac{\sigma_S}3$). Trade at $\mu_S-\nicefrac{\sigma_S}3$ forces a conditional mean shift of at least $\nicefrac{\sigma_S}3$, so the information identity controls these outliers' contribution to the second moment. Thus either conditioning on trade at $p$ substantially changes the mean or spread, or the outliers needed to preserve the spread provide information at $\mu_S-\nicefrac{\sigma_S}3$. Combining the bounds from these two prices gives
\[
 \alpha(p)\le\frac{225\overline D}{2\sigma_S}. \tag{Close-to-mean Bound}
\]
The buyer's bound is symmetric.
For values of $p$ close to the average of $\mu_S$ and $\mu_B$, we employ the trivial upper bound $\alpha(p) \le 1.$ Integrating $H(p)$ over $\R$ provides the desired upper bound on $G$ (and thus on $L(p)$) in terms of $\overline D$. Integrating the tail bounds {yields} $6\overline D$, while the close-to-mean regions have width $2\sigma_Z/3$ and height $225\overline D/(2\sigma_Z)$, for a total area of $162\overline D$. Finally, the central plateau has height one and length $(m-\nicefrac{\sigma}3)_+$, hence
\[
 G\le\int_{\R}H(p)\,dp
 \le162\overline D+(m-\nicefrac{\sigma}3)_+.
\]
Since $m\le G$, this implies $\min\{G,\nicefrac{\sigma}3\}\le162\overline D$. Also, Cauchy--Schwarz gives $G\le m_++\sigma<4\sigma$, so
\[
 G\le12\min\{G,\nicefrac{\sigma}3\}\le1944\overline D.
\]
If $\overline D=0$, then $G=0$ and any price works. Otherwise, since $1944<2^{11}$, a price sufficiently close to maximizing $D$ satisfies $L(p)\le G<2^{11}D(p)$.

\emph{Well-separated case ($m\ge3\sigma$).} In this case, {the seller's and buyer's means are sufficiently far apart that we can post their midpoint as the price.} A failed trade forces
either
$S-\mu_S>\nicefrac{m}2$ or $\mu_B-B>\nicefrac{m}2$; thus, conditional on rejection, the combined root-mean-square deviation is at least $\nicefrac{m}2\ge\nicefrac{3\sigma}2$, compared with at most $\sigma$ before conditioning. This increase requires changes in the conditional means or standard deviations. Their probability-weighted squared changes are bounded by $2\sigma D(p)$, while the missed gain is controlled by the same valuation deviations. Comparing these bounds gives
$L(p)\le41D(p)$.
\end{proof}

    {Our Bayesian algorithm posts the price supplied by the lemma.} 

\begin{algorithm}[H]
\caption{Bayesian Contextual Bilateral Trade}\label{alg:trade-bayesian}
\vspace{3pt}
\textbf{Input:} a known joint prior $\pi$ on $(\bs,\bb)$ with finite second moments.

Initialize the posterior $\pi_1 = \pi$.

For $t=1,2,\ldots$:
\begin{enumerate}
 \item Receive $\bx_t$ and let $\pi_t$ be the current posterior of $(\bs,\bb)$.
 \item Apply Lemma~\ref{lem:toi} under $\pi_t$ to
 $S=\bx_t^\top \bs$ and $B=\bx_t^\top \bb$ and post the resulting price $p_t$.
 \item Observe $Y_t$ and condition the joint posterior on this bit and the posted offer.
\end{enumerate}
\end{algorithm}

\begin{theorem}[Bayesian Bilateral Trade]\label{thm:trade-bayes}
For every known joint prior $\pi$ with finite second moments, Algorithm~\ref{alg:trade-bayesian} satisfies {the following bound at every horizon:}
\begin{equation}\label{eq:trade-bayesian-guarantee}
 \sum_{t=1}^T \E{ \ell_t}\le2^{11}\Psi_{\mathrm{tr}}(\pi).
\end{equation}
\end{theorem}

\begin{proof} Condition on the history and current context, and let $\pi_t^+$ be the posterior after the offer and trade bit. With $D(p_t)$ evaluated under $\pi_t$, the price choice and Lemma~\ref{lem:directional-sd-decrease} give
\begin{equation}\label{eq:trade-round}
 \Esub{t}{\ell_t}
 \overset{\text{Lemma~\ref{lem:toi}}}{\le}2^{11}D(p_t)
 \overset{\text{Lemma~\ref{lem:directional-sd-decrease}}}{\le}
 2^{11}\bigl(\Psi_{\mathrm{tr}}(\pi_t)-\Et\Psi_{\mathrm{tr}}(\pi_t^+)\bigr).
\end{equation}
The second inequality applies the directional comparison of Lemma~\ref{lem:directional-sd-decrease} to each marginal. It holds for any observed signal, including a trade bit that depends on both agents. By the context assumption in Section~\ref{sec:notation}, the next context supplies no additional information about the hidden pair. Taking expectations and telescoping therefore yields
\[
 \sum_{t=1}^T\E{\ell_t}
 \le2^{11}\bigl(\Psi_{\mathrm{tr}}(\pi)-\E{\Psi_{\mathrm{tr}}(\pi_T^+)}\bigr)
 \le2^{11}\Psi_{\mathrm{tr}}(\pi).
\]
The common price ensures per-round budget balance. The normalized bounds follow from~\eqref{eq:trade-bayesian-potential}.
\end{proof}

\subsection{Adversarial Contextual Bilateral Trade}\label{sec:trade-direct}

{In the adversarial setting, no prior is given, and the hidden pair is fixed:} $\boldsymbol w=(\boldsymbol s,\boldsymbol b)\in K=\mathbb B_2^d\times\mathbb B_2^d$. Our algorithm maintains a center $\bm=(\bm_S,\bm_B)\in\R^{2d}$ and a block-diagonal positive-definite matrix $R=\operatorname{diag}(R_S,R_B) \in \R^{2d \times 2d}$, where both $R_S$ and $R_B$ {belong} to $\R^{d\times d}$. These three quantities induce the same potential as in search, but {in dimension} $2d$:
\begin{equation}\label{eq:trade-quadratic}
 \Phi(\bm,R,\boldsymbol w)
 =\Tr R+(\boldsymbol w-\bm)^\top R^{-1}(\boldsymbol w-\bm)
 =\Phi(\bm_S,R_S,\boldsymbol s)+\Phi(\bm_B,R_B,\boldsymbol b).
\end{equation}
Note that the learner never evaluates the potential at the unknown pair $\boldsymbol{w}$. In each time step, our algorithm designs a suitable \emph{plan}, which specifies an offer $p$ and two successor states $(\bm^{(0)},R^{(0)})$ and $(\bm^{(1)},R^{(1)})$, one for each possible feedback bit. {Let} $\bm^+,R^+$ {denote }the random successor state induced by the {observed feedback}. Ideally, we would like to find a plan that balances \emph{decreasing} the expected potential at $(\bm^+,R^+,\boldsymbol{w})$ and {incurring little regret} $\ell(\boldsymbol{w})$. The following lemma {shows} that there exists a lottery over prices (i.e., a randomized plan) {that ensures this property}, up to some slack.

\begin{restatable}[A valid lottery exists]{lemma}{tradelottery}\label{lem:trade-lottery}
For every $\bm\in\R^{2d}$, every block-diagonal $R\succ0$, every context $\bx$ with $\norm{\bx}\le1$, and every $\gamma\in(0,1]$, there is a finite lottery of plans satisfying
\begin{equation}\label{eq:trade-pointwise}
 \E{\ell(\boldsymbol w)+2^{10}\Phi(\bm^+,R^+,\boldsymbol w)}
 \le2^{10}\Phi(\bm,R,\boldsymbol w)+\gamma
 \qquad\text{for every }\boldsymbol w\in K.
\end{equation}
{Each plan in the lottery posts the same price, drawn from a finite grid, to both agents. The lottery can be chosen as a function of the state and context.}
\end{restatable}
{Note that} the expectation in the lemma is only over the lottery, and the successor is selected using the bit generated by {the sampled plan}.
For a given slack $\gamma\in(0,1]$, the lottery must satisfy the one-round bound in \eqref{eq:trade-pointwise}, {which balances incurred regret against the decrease in the (unknown) potential and must hold for all possible hidden pairs.} We provide here a proof sketch of the lottery lemma, and defer the detailed proof to Appendix~\ref{sec:trade-lottery-proof}.

\begin{proof}[Proof sketch] The goal is to show that, for every fixed current state and context, we can first construct a good \emph{plan} {for} each hypothetical prior, and then use minimax to obtain one \emph{lottery} that works for every hidden pair. A \emph{plan} $a$ consists of a common price and two successor states, one for rejection and one for trade, all fixed before observing the bit. A \emph{lottery} $Q$ is a probability distribution over {plans:} the learner samples a plan, posts its price, and keeps the successor selected by the observed bit. We seek to find such a lottery and to ensure that it has finite support.

Let us write $\Phi_0(\boldsymbol w)=\Phi(\bm,R,\boldsymbol w)$ for the current quadratic potential from~\eqref{eq:variational-quadratic-Phi}, and $\Phi_a^+(\boldsymbol w)$ for the successor quadratic potential selected when the hidden pair is $\boldsymbol w$. Moreover, let us define the excess cost
\[
    f_a(\boldsymbol w)=\ell_a(\boldsymbol w)+2^{10}\Phi_a^+(\boldsymbol w)-2^{10}\Phi_0(\boldsymbol w), \tag{Excess cost}
\]
where $\ell_a$ is the lost gain under plan $a$. The required inequality is therefore $\Esub{a\sim Q}{f_a(\boldsymbol w)}\le\gamma$ for every $\boldsymbol w\in K$. Indeed, $f_a$ is the loss plus the scaled successor potential minus the scaled current potential, so this is exactly~\eqref{eq:trade-pointwise}.

\emph{A plan for each prior.} Temporarily suppose that the hidden pair $W$ is sampled from a known prior $\pi$ on $K$. Lemma~\ref{lem:toi}, together with the directional comparison of Lemma~\ref{lem:directional-sd-decrease}, supplies a common price $p$ satisfying 
\[ L(p)\le2^{11}\left(\Psi_{\mathrm{tr}}(\pi)-\mathbb P_\pi(Y(p)=0)\Psi_{\mathrm{tr}}(\pi_0)-\mathbb P_\pi(Y(p)=1)\Psi_{\mathrm{tr}}(\pi_1)\right), \tag{Bayesian bound} \] where $\pi_y=\mathcal L_\pi(W\mid Y(p)=y)$ is the posterior after observing bit $y$.
Thus the scaled prior potential covers both the expected loss and the scaled posterior potential remaining after the bit. We want to replace these potentials by the quadratic potentials of the current and successor states.

For each bit, choose the posterior mean as the successor center and the square roots of {the posterior} marginal covariances as the matrix blocks, adding $\lambda I_d$ before taking the roots to ensure positive definiteness. These choices define a plan $a_\lambda$. We stress that both successors are fixed before posting the price, and {the observed bit only selects which successor is kept}. The variational identity~\eqref{eq:variational-quadratic-Phi} gives \[ \begin{aligned} \Esub{\pi}{\Phi_0(W)}&\ge2\Psi_{\mathrm{tr}}(\pi),\\ \lim_{\lambda\to0^+}\Esub{\pi}{\Phi_{a_\lambda}^+(W)}&=2\left(\mathbb P_\pi(Y(p)=0)\Psi_{\mathrm{tr}}(\pi_0)+\mathbb P_\pi(Y(p)=1)\Psi_{\mathrm{tr}}(\pi_1)\right). \end{aligned} \tag{Potential bounds} \] 
Combining these comparisons with the Bayesian bound cancels the posterior terms: \[ \begin{aligned} \inf_a\Esub{\pi}{ f_a(W)}&\le\lim_{\lambda\to0^+}\Esub{\pi}{f_{a_\lambda}(W)}\\ &\le2^{11}\Psi_{\mathrm{tr}}(\pi)-2^{10}\Esub{\pi}{\Phi_0(W)}\le0. \end{aligned} \tag{Average bound} \] We have therefore constructed plans with arbitrarily small average excess cost for every prior, although the plans may depend on that prior.

\emph{One lottery for every hidden pair.} It remains to turn a suitable plan for \emph{each} prior into a \emph{single} lottery that works for every hidden pair. The obstacle to applying minimax directly is that the trade bit jumps when a valuation crosses the posted price. In the analysis, we slightly contract the hidden pair toward the origin and add a small common random shift to both valuations. The contraction keeps the perturbed pair in $K$, while averaging over the shift makes the expected cost continuous. Corresponding changes to the price and successor states preserve the averaged bound up to a uniformly vanishing error. We then apply Claim~\ref{cl:trade-minimax} together with Sion's minimax theorem (Theorem~\ref{thm:sion-minimax}~\cite{Sion58}) to these continuous costs {and then let the perturbation vanish}, giving 
\[
    \inf_Q\sup_{\boldsymbol w\in K}\Esub{a\sim Q}{f_a(\boldsymbol w)}=\sup_{\rho\in\mathcal P(K)}\inf_a\Esub{\rho}{f_a(W)}\le0. \tag{Minimax}
\]
Above, we write $\mathcal P(K)$ for the set of all probability measures on $K$, representing the possible prior distributions of the hidden pair. At this point, $Q$ ranges over arbitrary lotteries, not necessarily with finite support. Since the minimax value is at most zero, we can find a lottery with excess cost below any positive tolerance. We choose $\gamma/3$, leaving the remaining $2\gamma/3$ for the price and successor-state approximations below. Consequently, there is a lottery satisfying
\[
    \sup_{\boldsymbol w\in K}\Esub{a\sim Q}{f_a(\boldsymbol w)}<\frac{\gamma}{3}. \tag{Uniform bound}
\]
This bound holds for every fixed hidden pair, with expectation only over the sampled plan. 

\emph{A finite lottery.} The same claim lets us choose $Q$ with a bounded price density and compact ranges of prices and successor states. Using this regularity and compactness, we round prices to a finite grid and approximate successor states by a finite net, making each error smaller than $\gamma/3$. Thus, the resulting lottery has finite support and satisfies the required bound with slack $\gamma$. 

The detailed proof is deferred to Appendix~\ref{sec:trade-lottery-proof}. \end{proof}

\begin{algorithm}[H]
\caption{Adversarial Contextual Bilateral Trade}\label{alg:trade-adversarial}
\vspace{3pt}
Initialize $\bm_1=0\in\R^{2d}$ and $R_1=I_{2d}/\sqrt d$.

For $t=1,2,\ldots$:
\begin{enumerate}
 \item Receive $\bx_t$ and set $\gamma_t=\tfrac{1}{t(t+1)}$.
 \item Choose a {lottery of plans} as in Lemma~\ref{lem:trade-lottery} with current state $(\bm_t,R_t)$, context $\bx_t$, and slack $\gamma_t$.
 \item Draw a plan from the finite lottery and post its price $p_t$. 
 \item Observe $Y_t$ and retain the plan's successor state for that bit as $(\bm_{t+1},R_{t+1})$.
\end{enumerate}
\end{algorithm}

The algorithm uses only its state, the current context, its own randomization, and the observed bit. It needs neither a prior nor a horizon. {We do} not claim an efficient implementation of the optimization in step 2.

\begin{theorem}[Adversarial Bilateral Trade]\label{thm:trade-direct}
For every fixed $\boldsymbol w\in K$, every admissible adaptive context strategy with $\norm{\bx_t}\le1$, and every horizon $T$, Algorithm~\ref{alg:trade-adversarial} satisfies
\begin{equation}\label{eq:trade-direct-guarantee}
 \sum_{t=1}^T\E{\ell_t}\le2^{12}\sqrt d+1.
\end{equation}
\end{theorem}

\begin{proof}

Fix $\boldsymbol w=(\boldsymbol s,\boldsymbol b)\in K$ and write
$\Phi_t=\Phi(\bm_t,R_t,\boldsymbol w)$. Lemma~\ref{lem:trade-lottery}
guarantees the existence of the lottery required by Algorithm~\ref{alg:trade-adversarial}
at every round. Conditional on the observed history and current context, before
the lottery is drawn,~\eqref{eq:trade-pointwise} gives
\[
 \Esub{t}{\ell_t}+2^{10}\Esub{t}{\Phi_{t+1}}
 \le2^{10}\Phi_t+\gamma_t.
\]
Since $\Phi_1$ is finite and all losses and potentials are nonnegative,
this inequality also establishes their integrability inductively. Taking full
expectations and summing over $t=1,\ldots,T$ telescopes the potential terms:
\[
 \sum_{t=1}^T\E{\ell_t}+2^{10}\E{\Phi_{T+1}}
 \le2^{10}\Phi_1+\sum_{t=1}^T\gamma_t.
\]
The chosen slacks have total sum at most one, because
\[
 \sum_{t=1}^T\gamma_t
 =\sum_{t=1}^T\frac1{t(t+1)}
 =\sum_{t=1}^T\left(\frac1t-\frac1{t+1}\right)
 =1-\frac1{T+1}<1.
\]
Finally, the initialization and $\norm{\boldsymbol s},\norm{\boldsymbol b}\le1$ give
\[
 \Phi_1=\Phi(0,I_{2d}/\sqrt d,\boldsymbol w)
 =2\sqrt d+\sqrt d(\norm{\boldsymbol s}^2+\norm{\boldsymbol b}^2)\le4\sqrt d.
\]
Substituting these two bounds proves~\eqref{eq:trade-direct-guarantee}.
\end{proof}

\section*{Acknowledgment}

    Federico Fusco, Stefano Leonardi, and Matteo Russo acknowledge OpenAI for providing complimentary access through the ChatGPT for Academic Researchers program. The work of Federico Fusco and Stefano Leonardi is partly supported by AI-PHOQUS PN RIC 2021-2027, co-financed by EU Italian Ministry MIMIT and MUR. Anupam Gupta was supported in part by NSF awards CCF-2422926 and CCF-2608359.

\section*{Disclosure of AI Usage}

The first proof of $O(\sqrt{d})$ regret for the adversarial version of contextual bilateral trade was suggested by {GPT-6 Astra in} conversations with the authors. Subsequently, we dedicated considerable effort to simplifying the proofs, to {understanding and interpreting} the potential $\Phi$ used in the proof, to exploring the (somewhat simpler) setting of contextual search, to {identifying concrete connections} between the Bayesian and adversarial versions of both problems, and to investigating a Gaussian setting of contextual search, which led to interpreting the parameters of the algorithm as posterior mean and covariance proxies. 
The authors have {substantially rewritten the proofs} (albeit with some help from Gemini, {Claude, and ChatGPT) and take} full responsibility for the content of this work.

All the proofs have been formalized in Lean4 and can be found at \href{http://github.com/renatoppl/trace-square-root}{this url}.

\clearpage
\bibliographystyle{alpha}
\bibliography{references}

\clearpage

\appendix 
{\noindent \LARGE \textbf{Appendix}}
\section{Notation}\label{ap:nota} 

We summarize the main notations used in the paper in the table below. We generally use bold symbols for $d$-dimensional vectors, capital Roman letters for matrices and scalar random variables, and Greek letters for distributional quantities (such as moments) and potentials. Time dependence is usually indicated by a subscript; for example, $\Psi_t=\Psi(\Sigma_t)$ denotes the potential at the $t$-th round.

\begin{center}
\begin{tabular}{lccc}
  Quantity & Dimension $d$ & Projection & Random variable \\
  \hline
    \noalign{\vskip 6pt}
    \multicolumn{4}{l}{\emph{~~~~Contextual search}}\\[4pt]
  Unknown target & $\bz$ & $z$ & $\bz, Z$ \\
  Distributional mean & $\E{\bz} = \bmu$ & $\mu$ & \\
  Mean proxy & $\bm$ & $m$ & \\
  Distributional covariance & $\Cov(\bz) = \Sigma$ & $\sigma^2$ & \\
  Square root of covariance & $\sqrt{\Cov(\bz)} =\Sigma^{1/2}$ & $\sd(Z) = \sigma$ & \\
  Square root covariance proxy & $R$ & $r$ & \\
  Context & $\bx$ & & \\
  Price & & $p\in \Rbb$ & $P\in\mathbb{R}$ \\
  Feedback & & $y\in\{-1,1\}$ & $Y\in\{-1,1\}$ \\
  Random sign (Rademacher) & & & $\varepsilon\in\{-1,1\}$\\[6pt]
  \multicolumn{4}{l}{\emph{~~~~Contextual bilateral trade}}\\[4pt]
  Unknown seller/buyer valuations
    & $\boldsymbol w = (\boldsymbol s,\boldsymbol b)$ & ${S,B}$& $S,B \in \Rbb$ \\
    Joint law &$\pi = \mathcal L(\bs,\bb)$\\[6pt]
  \multicolumn{4}{l}{\emph{~~~~Potentials and functions}}\\[4pt]
  Bayesian potential & \multicolumn{3}{l}{$\Psi(\Sigma) = \Tr\sqrt{\Sigma}$}\\
  Bayesian trade potential
  & \multicolumn{3}{l}{$\Psi_{\mathrm{tr}}(\pi) = \Psi(\Cov(\bs))+\Psi(\Cov(\bb))$} \\
  Worst-case potential &  \multicolumn{3}{l}{$\Phi(\bm,R,\bz) = \Tr R+(\bz-\bm)^\top R^{-1}(\bz-\bm)$} \\
  Scalar potential & \multicolumn{3}{l}{$\varphi(m,r,z) = r+(m-z)^2/r$} \\
  Gap function & \multicolumn{3}{l}{$g_{\bx}(\Sigma) = \Tr\sqrt{\Sigma}-\sqrt{\bx^\top\Sigma\bx}$ }\\[4pt]
  \hline
\end{tabular}
\end{center}
\vspace{10pt}

\section{Concavity of the Gap Function}\label{app:gap-concavity}

This appendix proves the concavity and monotonicity of the gap function $g_{\bx}$ used in Lemma~\ref{lem:directional-sd-decrease}. We do so by expressing $g_{\bx}$ as an infimum of affine functions with positive-semidefinite slopes.

\begin{lemma}[Gap representation and directional comparison]\label{lem:envelope}
Fix $\bx\in\R^d$ with $\norm{\bx}\le1$ and define, for $\Sigma\succeq0$,
\begin{equation}\label{eq:bayesian-potential}
g_{\bx}(\Sigma)=\Tr\sqrt{\Sigma}-\sqrt{\bx^\top\Sigma\bx}.
\end{equation}
The function $g_{\bx}$ admits the affine representation
\begin{equation}\label{eq:envelopeexact}
 2g_{\bx}(\Sigma)=\inf_{R\succ0}\left\{
 \Tr R-\norm{R\bx}
 +\Tr\!\left[\left(R^{-1}-\frac{\bx\bx^\top}{\norm{R\bx}}\right)\Sigma\right]\right\}.
\end{equation}
Here $\bx\bx^\top/\norm{R\bx}$ is defined to be zero when $\bx=0$. If $\Sigma\succ0$, the infimum is attained at $R=\Sigma^{1/2}$.

Each function inside the infimum in~\eqref{eq:envelopeexact} is affine in $\Sigma$ and has positive-semidefinite slope. Consequently, $g_{\bx}$ is concave as a pointwise infimum of affine functions, and it is {nondecreasing in the positive semidefinite order}. Applied to $\Et[\Sigma_{t+1}]\preceq\Sigma_t$  in the Bayesian search model, the lemma thus implies
\begin{equation}\label{eq:directional-comparison}
 \Psi_t-\Esub{t}{\Psi_{t+1}}\ge\sigma_t-\Esub{t}{\sigma_t'}.
\end{equation}
\end{lemma}

\begin{proof}[Proof of Lemma~\ref{lem:envelope}]
Fix $R\succ0$ and put $T=\Sigma^{1/2}$, $M=(T-R)^2\succeq0$, $r=\norm{R\bx}$, and $\sigma=\norm{T\bx}$. Expanding the square and using {cyclicity of the trace} gives
\begin{equation}\label{eq:envexpand}
 \Tr(R^{-1}M)=\Tr(R^{-1}\Sigma)-2\Tr T+\Tr R.
\end{equation}
For $\bx\ne0$, Cauchy--Schwarz in the $R$-inner product, followed by $\bx^\top R\bx\le\norm{\bx}\norm{R\bx}\le r$, yields
\[
 R^{-1}\succeq\frac{\bx\bx^\top}{\bx^\top R\bx}
 \succeq\frac{\bx\bx^\top}{r}.
\]
Thus $H_R=R^{-1}-\bx\bx^\top/r\succeq0$, and the reverse triangle inequality gives
\[
 \Tr(R^{-1}M)
 \ge\frac{\bx^\top M\bx}{r}
 =\frac{\norm{(T-R)\bx}^2}{r}
 \ge\frac{(\sigma-r)^2}{r}.
\]
{Combining this inequality with}~\eqref{eq:envexpand} and rearranging proves
\begin{equation}\label{eq:envelope}
 2\Tr T-2\sigma\le\Tr R-r+\Tr(H_R\Sigma).
\end{equation}
For $\bx=0$, take $H_R=R^{-1}$; the same bound follows directly from~\eqref{eq:envexpand}. If $\Sigma\succ0$, equality holds at $R=T$. For singular $\Sigma$, the same value is approached by $R_\eta=(\Sigma+\eta I)^{1/2}$ as $\eta\to 0^+$.
\end{proof}

\section{Supporting Proofs for Contextual Search}\label{app:proofs}

We have deferred two proofs in \Cref{sec:search}: Lemma~\ref{lem:deviation-contraction} in the Bayesian analysis and Lemma~\ref{lem:drift} in the worst-case analysis. Both rely on the same one-dimensional potential decrease, stated in Lemma~\ref{lem:scalar-potential} below. We first establish this common inequality, then use it to complete the two proofs.

\subsection{A One-Dimensional Potential Lemma}\label{app:scalar}

\begin{lemma}[One-dimensional potential decrease]\label{lem:scalar-potential}
For $m,z\in\R$ and $r>0$, define
\[
 \varphi(m,r,z)=r+\frac{(m-z)^2}{r}.
\]
Let $\varepsilon$ be uniform on $\{-1,1\}$, set
\[
 P=m+\frac r2\varepsilon,\quad
 Y=\operatorname{sign}(z-P),
\]
and define
\[
 m^+=m+\frac{rY}{3-\varepsilon Y},\qquad r^+=\frac{3r}{3-\varepsilon Y}.
\]
Then
\begin{equation}\label{eq:scalar-potential-drift}
 \varphi(m,r,z)-\Esub{\varepsilon}{\varphi(m^+,r^+,z)}
 \ge\frac16\Esub{\varepsilon}{|z-P|}
 =\frac16\max\left\{|z-m|,\frac r2\right\}.
\end{equation}
In particular,
\begin{equation}\label{eq:scalar-scale-decrease}
 \Esub{\varepsilon}{\varphi(m^+,r^+,z)}\le\varphi(m,r,z)-\frac r{12}.
\end{equation}
\end{lemma}

\begin{figure}[htbp]
\centering
\begin{tikzpicture}[
  x=5.3cm, y=2.2cm,
  line cap=round, line join=round, >=Stealth
]
\definecolor{scalarinterior}{RGB}{54,95,136}
\definecolor{scalarexterior}{RGB}{180,85,35}

\foreach \left/\right/\regioncolor in
  {-1.15/-0.5/scalarexterior,-0.5/0.5/scalarinterior,0.5/1.15/scalarexterior}{
  \path[fill=\regioncolor!20] (\left,0) rectangle (\right,0.36);
}
\node[text=scalarexterior] at (-0.83,0.18) {Exterior};
\node[text=scalarinterior] at (0,0.18) {Interior};
\node[text=scalarexterior] at (0.83,0.18) {Exterior};

\foreach \price in {-0.5,0.5}{
  \draw[scalarinterior!50,line width=0.5pt] (\price,0.39) -- (\price,0.61);
}
\draw[<->,scalarinterior,line width=0.8pt]
  (-0.5,0.54) -- node[above,inner sep=2pt] {$r$} (0.5,0.54);

\foreach \price in {-0.5,0.5}{
  \draw[black!45,dashed,line width=0.6pt] (\price,0) -- (\price,0.36);
}
\draw[<->,line width=0.8pt] (-1.22,0) -- (1.22,0) node[right] {$z$};
\foreach \point in {-0.5,-0.25,0,0.25,0.5}{
  \draw (\point,-0.025) -- (\point,0.025);
  \fill (\point,0) circle[radius=1.6pt];
}

\node[anchor=north] at (-0.5,-0.065) {$p_-=m-\frac r2$};
\node[anchor=north] at (0,-0.065) {$m$};
\node[anchor=north] at (0.5,-0.065) {$p_+=m+\frac r2$};

\foreach \center in {-0.25,0.25}{
  \draw[scalarinterior!55,line width=0.5pt]
    (\center,-0.035) -- (\center,-0.42);
}
\node[anchor=north] at (-0.25,-0.45) {$m-\frac r4$};
\node[anchor=north] at (0.25,-0.45) {$m+\frac r4$};
\end{tikzpicture}
\caption{Prices and possible updated centers $m^+\in\{m\pm r/4,\,m\pm r/2\}$ in
Lemma~\ref{lem:scalar-potential}.}
\label{fig:scalar-regions}
\end{figure}
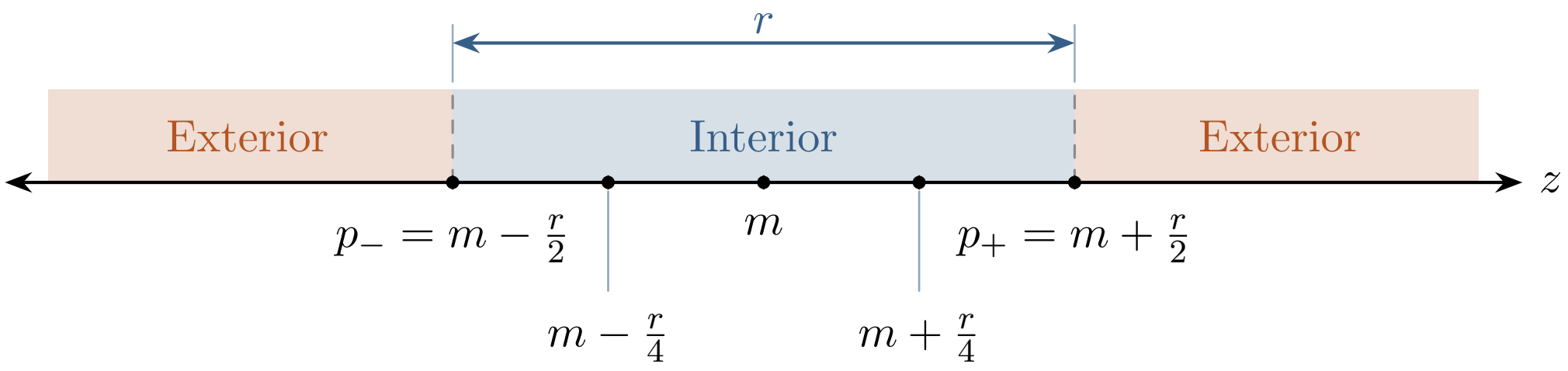

\begin{proof}
The identity for the expected loss follows by averaging over the two prices. We prove the potential bound by separating targets between the two prices from targets {outside that interval}, as illustrated in Figure~\ref{fig:scalar-regions}.

\paragraph{Interior: $|z-m|<r/2$.}
Both answers point inward, so $\varepsilon Y=-1$ and $r^+=3r/4$. The new center is $m+r/4$ or $m-r/4$, each with probability $1/2$. Thus
\[
 \Esub{\varepsilon}{(m^+-z)^2}=(m-z)^2+\frac{r^2}{16},
\]
and
\begin{align*}
 \Esub{\varepsilon}{\varphi(m^+,r^+,z)}
 &=\frac{3r}{4}+\frac4{3r}\left((m-z)^2+\frac{r^2}{16}\right)\\
 &=\varphi(m,r,z)-\frac r4+\frac{(m-z)^2}{3r}+\frac r{12}
 \le\varphi(m,r,z)-\frac r{12}.
\end{align*}
Here $\Esub{\varepsilon}{|z-P|}=\nicefrac{r}2$, proving~\eqref{eq:scalar-potential-drift} in this case.

\paragraph{Exterior: $|z-m|>\nicefrac{r}2$.}
{By reflection,} it suffices to consider $z>m+\nicefrac{r}2$. Both prices {yield the answer} $Y=1$, and the two equally likely successor states are
\[
 (m^+,r^+)=(m+r/4,3r/4)
 \quad\text{and}\quad(m+r/2,3r/2).
\]
Consequently,
\[
 \Esub{\varepsilon}{r^+}=\frac{9r}{8},
\]
and
\begin{align*}
 \Esub{\varepsilon}{\frac{(z-m^+)^2}{r^+}}
 &=\frac12\frac{(z-m-r/4)^2}{3r/4}
   +\frac12\frac{(z-m-r/2)^2}{3r/2}\\
 &=\frac{(z-m)^2}{r}-\frac23(z-m)+\frac r8.
\end{align*}
Adding the two terms and using $z-m\ge r/2$ gives
\[
 \Esub{\varepsilon}{\varphi(m^+,r^+,z)}
 =\varphi(m,r,z)+\frac r4-\frac23(z-m)
 \le\varphi(m,r,z)-\frac16(z-m).
\]
Since $\Esub{\varepsilon}{|z-P|}=z-m$, this proves the claim for exterior targets.

\paragraph{Boundary: $|z-m|=r/2$.}
Our convention $\operatorname{sign}(0)=1$ gives one inward and one outward update when $z=m+r/2$, as in the exterior calculation. When $z=m-r/2$, both updates are inward, as in the interior calculation. Substituting into the corresponding expressions gives a decrease of exactly $r/12$ in either case, while the expected loss is $r/2$. This proves~\eqref{eq:scalar-potential-drift} at both endpoints. Finally, the expected loss is always at least $r/2$, which implies~\eqref{eq:scalar-scale-decrease}.
\end{proof}

\subsection{Proof of Lemma~\ref{lem:deviation-contraction}}\label{app:contraction}

\deviationcontraction*

\begin{proof}[Proof of Lemma~\ref{lem:deviation-contraction}]
If $\sigma=0$, then $Z=\mu$ almost surely and $\sigma'=0$, so the result is immediate. Assume $\sigma>0$. For any mean proxy $m\in\R$ and deviation proxy $r>0$,
\begin{equation}\label{eq:quadratic-bound}
 2\sd(Z)\le2\sqrt{\E{(Z-m)^2}}
 \le r+\frac{\E{(Z-m)^2}}r
 =\E{\varphi(m,r,Z)}.
\end{equation}
Equality holds at $m=\mu$ and $r=\sigma$. We apply this bound conditionally after observing the price and answer, with the proxies
\[
 m^+=\mu+\frac{\sigma Y}{3-\varepsilon Y},
 \qquad r^+=\frac{3\sigma}{3-\varepsilon Y}.
\]
Because $\sigma>0$, the sign $\varepsilon$ is determined by $P$, so these proxies are $(P,Y)$-measurable. Applying~\eqref{eq:quadratic-bound} to the conditional law yields
\[
 2\sigma'\le\E{\varphi(m^+,r^+,Z)\mid P,Y},
 \qquad
 2\E{\sigma'}\le\E{\varphi(m^+,r^+,Z)}.
\]
For each fixed $Z=z$, Lemma~\ref{lem:scalar-potential} with $m=\mu$ and $r=\sigma$ gives
\[
 \E{\varphi(m^+,r^+,Z)\mid Z=z}
 \le\varphi(\mu,\sigma,z)-\frac\sigma{12}.
\]
Averaging over $Z$ and using $\E{\varphi(\mu,\sigma,Z)}=2\sigma$ gives
\[
 2\E{\sigma'}\le2\sigma-\frac\sigma{12}=\frac{23}{12}\sigma.
\]
Division by two proves the lemma.
\end{proof}

\subsection{Proof of Lemma~\ref{lem:drift}}\label{app:drift}

\potentialdecrease*

\begin{proof}[Proof of Lemma~\ref{lem:drift}]
For $\bx_t=0$, the loss and potential change are both zero. Otherwise, condition on the history and current context, suppress time subscripts, and put
\[
 m=\bx^\top\bm,\quad z=\bx^\top\bz,\quad
 \boldsymbol u=R\bx,\quad a=\bx^\top R\bx,\quad r=\norm{R\bx}.
\]
Since $R\succ0$ and $\norm{\bx}\le1$, we have
\begin{equation}\label{eq:a-le-r}
 0<a=\bx^\top\boldsymbol u\le\norm{\bx}\norm{\boldsymbol u}\le r.
\end{equation}
{For a realization of the random sign and the answer}, let $\bm_+,R_+$ be the updated state. The update formulas imply
\[
 m^+=\bx^\top\bm_+=m+\frac{Yr}{3-\varepsilon Y},
 \qquad R_+\bx=\frac3{3-\varepsilon Y}R\bx,
 \qquad r^+=\norm{R_+\bx}=\frac{3r}{3-\varepsilon Y}.
\]
Thus $m^+,r^+$ are exactly the successor quantities in Lemma~\ref{lem:scalar-potential}.

We first prove the identity
\begin{equation}\label{eq:matrix-scalar-potential-change}
 \Phi(\bm_+,R_+,\bz)-\Phi(\bm,R,\bz)
 =\frac ra\bigl(\varphi(m^+,r^+,z)-\varphi(m,r,z)\bigr).
\end{equation}
The trace change is
\begin{equation}\label{eq:trace-change}
 \Tr(R_+-R)=\frac{\varepsilon Y r^2}{a(3-\varepsilon Y)}=\frac ra(r^+-r).
\end{equation}
For the quadratic term, write $\delta=m^+-m=Yr/(3-\varepsilon Y)$. The updates give
\[
 \bm_+-\bm=\frac\delta a\boldsymbol u,
 \qquad R_+^{-1}=R^{-1}-\frac{\varepsilon Y}{3a}\bx\bx^\top.
\]
The inverse identity follows directly by multiplication with $R_+$. Using $R^{-1}\boldsymbol u=\bx$ and $\boldsymbol u^\top R^{-1}\boldsymbol u=a$, we obtain
\begin{align*}
 &(\bz-\bm_+)^\top R_+^{-1}(\bz-\bm_+)\\
 &\quad=(\bz-\bm)^\top R^{-1}(\bz-\bm)
       -\frac{2\delta(z-m)}a+\frac{\delta^2}a
       -\frac{\varepsilon Y}{3a}(z-m^+)^2\\
 &\quad=(\bz-\bm)^\top R^{-1}(\bz-\bm)
       +\frac{3-\varepsilon Y}{3a}(z-m^+)^2-\frac{(z-m)^2}a\\
 &\quad=(\bz-\bm)^\top R^{-1}(\bz-\bm)
       +\frac ra\left(\frac{(z-m^+)^2}{r^+}-\frac{(z-m)^2}r\right).
\end{align*}
The second equality uses $z-m^+=(z-m)-\delta$; the third uses $r^+=3r/(3-\varepsilon Y)$. Adding~\eqref{eq:trace-change} proves~\eqref{eq:matrix-scalar-potential-change}.

Finally, take the conditional expectation and apply Lemma~\ref{lem:scalar-potential}:
\begin{align*}
 \Phi-\Esub{t}{\Phi_+}
 &=\frac ra\bigl(\varphi(m,r,z)-\Esub{t}{\varphi(m^+,r^+,z)}\bigr)\\
 &\ge\frac r{6a}\Esub{t}{|z-P|}
 \ge\frac16\Esub{t}{\ell_t},
\end{align*}
where the last inequality is~\eqref{eq:a-le-r}. This is the desired drift bound.
\end{proof}

\section{Supporting Proofs for Bilateral Trade}\label{sec:trade-proofs}

We first prove the trade-or-information inequality of Lemma~\ref{lem:toi}, which directly yields the Bayesian guarantee. We then use it, together with quadratic bounds derived from Lemma~\ref{lem:envelope} and a minimax argument, to prove the lottery-existence statement of Lemma~\ref{lem:trade-lottery}.

\subsection{Proof of Lemma~\ref{lem:toi}}\label{sec:trade-toi}

Recall that $Y(p)=\mathbf1\{S\le p\le B\}$ indicates whether trade occurs at price $p$. The expected lost gain and the expected reduction in the sum of the two standard deviations are, respectively,
\[
 \begin{aligned}
 L(p)&=\E{(B-S)_+(1-Y(p))},\\
 D(p)&=(\sd(S)-\E{\sd(S\mid Y(p))}) + (\sd(B)-\E{\sd(B\mid Y(p))}).
 \end{aligned}
\]

Before proving the lemma, {we show that} $D(p)$ is {nonnegative}, provide a useful {integral representation of the gain from trade}, and introduce a useful result on the information produced by binary feedback.

\begin{claim}
\label{cl:D_pos}
    For every pair $(S,B)$ of square-integrable real random variables,   $D(p) \ge 0$ for all $p \in \R$.
\end{claim}
\begin{proof}
    For each $Z\in\{S,B\}$, the law of total variance gives
\[
\operatorname{Var}(Z)
=
\mathbb{E}\bigl[\operatorname{Var}(Z\mid Y(p))\bigr]
+
\operatorname{Var}\bigl(\mathbb{E}[Z\mid Y(p)]\bigr)
\ge
\mathbb{E}\bigl[\operatorname{Var}(Z\mid Y(p))\bigr].
\]
Consequently, by Jensen's inequality and the concavity of the
square root,
\[
\begin{aligned}
\sum_{{y \in \{0,1\}}}\mathbb{P}(Y(p)=y)\operatorname{sd}(Z\mid Y(p)=y)
&=
\mathbb{E}\left[\sqrt{\operatorname{Var}(Z\mid Y(p))}\right]\\
&\le
\sqrt{\mathbb{E}\bigl[\operatorname{Var}(Z\mid Y(p))\bigr]}\\
&\le \operatorname{sd}(Z).
\end{aligned}
\]
Thus, assuming finite variances, each summand in the definition
of $D(p)$ is nonnegative, and hence $D(p)\ge 0$.
Here, outcomes $y$ with $\mathbb{P}(Y(p)=y)=0$ are omitted
from the sum.
\end{proof}

\paragraph{Trade profile and gain from trade.} {Here we give an integral representation} of the optimal expected gain from trade under $(S,B)$, {in terms of} the trade-profile function $\alpha(p) = \P{p \in [S,B]}.$ Denote the optimal expected gain from trade {by} $G=\E{(B-S)_+}$, where $(\cdot)_+$ denotes the positive part. {The claim} does not require independence and {also holds}
when both sides are infinite.

\begin{claim}[Gain-from-trade decomposition]
\label{cl:decomposition}
The following integral decomposition {holds}:
\[
G=\int_{\mathbb{R}}\alpha(p)\,dp.
\]
\end{claim}

\begin{proof}
For every realization $(s,b)$ of $(S,B)$, we can write the corresponding gain from trade as follows:
\[
(B-S)^+
=\int_{\mathbb{R}}\mathbf{1}\{S\le p\le B\}\,dp,
\]
since the integral is the length of the interval of prices
at which trade is possible.
Because the integrand is nonnegative, Tonelli's theorem gives
\[
    \mathbb{E}[(B-S)^+]
=\mathbb{E}\left[
    \int_{\mathbb{R}}\mathbf{1}\{S\le p\le B\}\,dp
\right]=\int_{\mathbb{R}}\mathbb{P}(S\le p\le B)\,dp
=\int_{\mathbb{R}}\alpha(p)\,dp. \qedhere
\]
\end{proof}

\paragraph{Information as displacement.} For any event $E$ of positive probability, we introduce two notions of {``distance''} to measure the information provided by the {occurrence} of $E$ {about} $S$ and $B$. 
\[
 \delta_{Z\mid E}^2=({\E{Z\mid E}}-\mu_Z)^2+(\sd(Z\mid E)-\sigma_Z)^2,
 \qquad
 \varrho_{Z\mid E}^2={\E{(Z-\mu_Z)^2\mid E}}.
\]
The first quantity $\delta_{Z\mid E}^2$ measures the distance from the posterior point $({\E{Z\mid E}},\sd(Z\mid E))$ to $(\mu_Z,\sigma_Z)$, while the second one $\varrho_{Z\mid E}^2$ is the conditional second moment around the unconditional mean \(\mu_Z\).

\begin{claim}[Information is displacement]\label{cl:trade-displacement}
For a square-integrable $Z$ and a binary signal $Y$,
\begin{equation}\label{eq:trade-displacement}
 \sum_{{y \in \{0,1\}}}\mathbb P(Y=y)\delta_{Z\mid\{Y=y\}}^2
 =2\sigma_Z\left(\sigma_Z-\sum_{{y \in \{0,1\}}}\mathbb P(Y=y)\sd(Z\mid Y=y)\right).
\end{equation}
For either positive-probability branch $E$, we also have
\begin{equation}\label{eq:trade-displacement-readings}
 |\E{Z\mid E}-\mu_Z|\le\delta_{Z\mid E},\qquad
 |\varrho_{Z\mid E}-\sigma_Z|\le\delta_{Z\mid E}.
\end{equation}
If $|\E{Z\mid E}-\mu_Z|\ge\eta>0$, then
$\varrho_{Z\mid E}\le(1+\sigma_Z/\eta)\delta_{Z\mid E}$.
\end{claim}
\begin{proof}
By the law of total variance,
\[
\sum_{{y \in \{0,1\}}} \mathbb{P}(Y=y)
\left(
    \bigl(\mathbb{E}[Z\mid Y=y]-\mu_Z\bigr)^2
    + \operatorname{sd}(Z\mid Y=y)^2
\right)
= \sigma_Z^2.
\]
Expanding the square in the definition of $\delta_{Z\mid\{Y=y\}}^2$
and using this identity gives
\[
\begin{aligned}
&\sum_{{y \in \{0,1\}}} \mathbb{P}(Y=y)\delta_{Z\mid\{Y=y\}}^2
\\
=~
&\sum_{{y \in \{0,1\}}} \mathbb{P}(Y=y)\!
\left[
    \bigl(\mathbb{E}[Z\mid Y=y]\!\!-\!\!\mu_Z\bigr)^2
   \!\! + \operatorname{sd}(Z\mid Y=y)^2
    \!\!-\! 2\sigma_Z\operatorname{sd}(Z\mid Y=y)
    + \sigma_Z^2
\right]\\
=~
&2\sigma_Z^2
-2\sigma_Z\sum_{{y \in \{0,1\}}}\mathbb{P}(Y=y)
    \operatorname{sd}(Z\mid Y=y)\\
=~
&2\sigma_Z
\left(
    \sigma_Z-\sum_{{y \in \{0,1\}}}\mathbb{P}(Y=y)
    \operatorname{sd}(Z\mid Y=y)
\right),
\end{aligned}
\]
which proves \eqref{eq:trade-displacement}.

{Fix an event $E$ of positive probability. The} first inequality in \eqref{eq:trade-displacement-readings} follows directly from
\[
\delta_{Z\mid E}^2
=
\bigl(\mathbb{E}[Z\mid E]-\mu_Z\bigr)^2
+
\bigl(\operatorname{sd}(Z\mid E)-\sigma_Z\bigr)^2
\ge
\bigl(\mathbb{E}[Z\mid E]-\mu_Z\bigr)^2.
\]
To prove the second inequality, first note that
\[
\begin{aligned}
\varrho_{Z\mid E}^2
&=\mathbb{E}\bigl[(Z-\mu_Z)^2\mid E\bigr]\\
&=\mathbb{E}\bigl[
    (Z-\mathbb{E}[Z\mid E])^2\mid E
\bigr]
+\bigl(\mathbb{E}[Z\mid E]-\mu_Z\bigr)^2\\
&=\operatorname{sd}(Z\mid E)^2
+\bigl(\mathbb{E}[Z\mid E]-\mu_Z\bigr)^2.
\end{aligned}
\]
Here the cross term vanishes because
$\mathbb{E}[Z-\mathbb{E}[Z\mid E]\mid E]=0$ and, in particular,
$\varrho_{Z\mid E}\ge\operatorname{sd}(Z\mid E)$.
Therefore,
\[
\begin{aligned}
\bigl(\varrho_{Z\mid E}-\sigma_Z\bigr)^2
&=\varrho_{Z\mid E}^2+\sigma_Z^2
    -2\sigma_Z\varrho_{Z\mid E}\\
&\le
\varrho_{Z\mid E}^2+\sigma_Z^2
    -2\sigma_Z\operatorname{sd}(Z\mid E)\\
&=
\bigl(\mathbb{E}[Z\mid E]-\mu_Z\bigr)^2
+\bigl(\operatorname{sd}(Z\mid E)-\sigma_Z\bigr)^2\\
&=\delta_{Z\mid E}^2.
\end{aligned}
\]
Taking square roots proves
$|\varrho_{Z\mid E}-\sigma_Z|\le\delta_{Z\mid E}$,
and hence \eqref{eq:trade-displacement-readings}.

Finally, if $|\mathbb{E}[Z\mid E]-\mu_Z|\ge\eta>0$,
the first inequality in \eqref{eq:trade-displacement-readings} gives $\delta_{Z\mid E}\ge\eta$.
Combining this with the second inequality yields
\[
\varrho_{Z\mid E}
\le \sigma_Z+\delta_{Z\mid E}
\le \frac{\sigma_Z}{\eta}\delta_{Z\mid E}
    +\delta_{Z\mid E}
=
\left(1+\frac{\sigma_Z}{\eta}\right)\delta_{Z\mid E}. \qedhere
\]
\end{proof}

We now proceed with the proof of Lemma~\ref{lem:toi}.

\tradeorinformation*

\begin{proof}
    We write $\mu_Z=\E{Z}$ and $\sigma_Z=\sd(Z)$ for $Z \in \{S,B\}$ {to denote the mean and standard deviation of valuation} $Z${, respectively}. As a {first-order} measure of the gain from trade achievable {for a given pair} $(S,B)${, we use} $m=\mu_B - \mu_S$, while we use $\sigma=\sigma_S+\sigma_B$ as a proxy for {the spread of the two variables}. We start by addressing some simple corner cases, and then move to the most challenging regime. 
    
    \paragraph{Deterministic values.} If either $\sigma_S$ or $\sigma_B$ {is} zero, then one of the two variables is deterministic and posting that value yields the optimal gain from trade. If $\sigma_{{S}}=0$, then posting $\mu_S$ is optimal and yields $L(\mu_S)=0 \le D(\mu_S)$ (by Claim~\ref{cl:D_pos}); similarly, posting $\mu_B$ when $\sigma_B=0$ yields the desired inequality. 

    \paragraph{Well-separated case.} This case is characterized by a large distance $m$ between the expected values of $S$ and $B$, {relative to their standard deviations}, i.e.,  $m\ge3\sigma$, and a price satisfying the statement of the lemma is $\overline  p=(\mu_S+\mu_B)/2$. Let $F=\{{Y(\overline p)}=0\}$ be the {event that trade fails at price} $\overline p$. If $\overline p$ always {results} in a trade, then Claim~\ref{cl:D_pos} yields the statement. {We may therefore assume} that $\mathbb P(F)>0$. Conditioning on $F$, at least one of $S-\mu_S$ and $\mu_B-B$ exceeds $m/2$. Hence
\[
 N=\sqrt{(S-\mu_S)^2+(B-\mu_B)^2}>m/2,
 \qquad (B-S)_+\le m+\sqrt2N\le(2+\sqrt2)N.
\]
Put $\nu=\sqrt{\E{N^2\mid F}}\ge m/2\ge3\sigma/2$. Then
\begin{equation}\label{eq:trade-large-gap-loss}
 L(\overline  p)\le(2+\sqrt2){\P{F}}\nu.
\end{equation}
Claim~\ref{cl:trade-displacement} applied to each agent under $F$ yields
\[
 \P{F}\bigl[(\varrho_{S\mid F}-\sigma_S)^2+(\varrho_{B\mid F}-\sigma_B)^2\bigr]
 \le2\sigma_SD_S(\overline p)+2\sigma_BD_B(\overline p)
 \le2\sigma D(\overline p).
\]
The vector $(\varrho_{S\mid F},\varrho_{B\mid F})$ has norm $\nu$, while $(\sigma_S,\sigma_B)$ has norm at most $\sigma$. The {quantity in brackets} is therefore at least $(\nu-\sigma)^2\ge\nu^2/9$. It follows that
\[
 D(\overline  p)\ge\frac{{\P{F}}\nu^2}{18\sigma}\ge\frac{{\P{F}}\nu}{12},
 \qquad L(\overline  p)\le12(2+\sqrt2)D(\overline  p)\le41D(\overline  p).
\]

\paragraph{Overlapping case.} We are left with the last case, i.e., $m < 3 \sigma$, which is arguably the most intricate: {the seller's and buyer's expected values} are relatively close, and we want to find a ``good'' price. We tackle this case by providing {a uniform} upper bound on the trade profile $\alpha(p)${, which} yields, by Claim~\ref{cl:decomposition}{,} the desired result. We refer to Figure~\ref{fig:trade-profile} for a visualization of this bound. {Define the supremum of $D$ on ${\R}$ by} $
 \overline D=\sup_{p\in\R}D(p).
$ {Note that} $0\le\overline D\le\sigma$. We first bound the trade profile uniformly over all prices {in terms of} the auxiliary {piecewise} function $H:$
\[
H(p)=
 \begin{cases}
  \dfrac{2\sigma_S\overline D}{(\mu_S-p)^2},
    & p<\mu_S-\sigma_S/3,\\[6pt]
  \dfrac{225\overline D}{2\sigma_S},
    & \mu_S-\sigma_S/3\le p\le\mu_S+\sigma_S/3,\\[6pt]
  1,
    & \mu_S+\sigma_S/3<p<\mu_B-\sigma_B/3,\\[6pt]
  \dfrac{225\overline D}{2\sigma_B},
    & \begin{aligned}
        &p>\mu_S+\sigma_S/3,\\
        &\mu_B-\sigma_B/3\le p\le\mu_B+\sigma_B/3,
      \end{aligned}\\[6pt]
  \dfrac{2\sigma_B\overline D}{(p-\mu_B)^2},
    & p>\max\{\mu_S+\sigma_S/3,\mu_B+\sigma_B/3\}.
 \end{cases}
\]
If the central intervals overlap, the seller's bound is used through $\mu_S+\sigma_S/3$ and the buyer's bound thereafter.
\begin{claim}[Uniform bound on the trade profile]\label{cl:trade-profile}
For every $p\in\R$, {we have} 
\(
\alpha(p)\le H(p)
 \){.}
\end{claim}
\begin{proof}
For any positive-probability branch $E$ of $Y(p)$, Claim~\ref{cl:trade-displacement} gives
\[
 \mathbb P(E)\delta_{Z\mid E}^2
 \le2\sigma_ZD_Z(p)\le2\sigma_Z\overline D,
 \qquad Z\in\{S,B\}.
\]
We prove the seller's tail and central bounds first. If $\alpha(p)=0$, both are immediate, so assume that $E=\{S\le p\le B\}$ has positive probability. For $p<\mu_S$, we have $\E{S\mid E}\le p$, and hence
\[
 \alpha(p)(\mu_S-p)^2
 \le\alpha(p)\delta_{S\mid E}^2
 \le2\sigma_S\overline D.
\]
This proves the upper bound in the first piece of $H$.

For $\mu_S-\sigma_S/3\le p\le\mu_S+\sigma_S/3$, set $\eta=\sigma_S/3$, $\alpha=\alpha(p)$, and $\varrho=\varrho_{S\mid E}$. Sellers trading at $p$ with $S\ge\mu_S-\eta$ are within $\eta$ of their mean. Those with $S<\mu_S-\eta$ also trade at the lower price $\mu_S-\eta$, since $B\ge p\ge\mu_S-\eta$. Thus, with $E'=\{S\le\mu_S-\eta\le B\}$,
\begin{equation}\label{eq:trade-near-split}
 \alpha\varrho^2
 \le\eta^2\alpha+\E{(S-\mu_S)^2\mathbf1_{E'}}.
\end{equation}
If $\mathbb P(E')>0$, then $\E{S\mid E'}\le\mu_S-\eta$. By Claim~\ref{cl:trade-displacement},
\[
 \mathbb P(E')\varrho_{S\mid E'}^2
 \le16\mathbb P(E')\delta_{S\mid E'}^2
 \le32\sigma_S\overline D.
\]
The same bound on the last expectation in~\eqref{eq:trade-near-split} is immediate if $\mathbb P(E')=0$. On $E$ itself, Claim~\ref{cl:trade-displacement} also gives
$\alpha(\sigma_S-\varrho)^2\le2\sigma_S\overline D$. Consequently,
\[
 \sigma_S\sqrt\alpha
 \le\sqrt{\eta^2\alpha+32\sigma_S\overline D}
       +\sqrt{2\sigma_S\overline D}
 \le\frac{\sigma_S}{3}\sqrt\alpha
       +5\sqrt{2\sigma_S\overline D}.
\]
Rearranging and squaring yields
$\alpha(p)\le225\overline D/(2\sigma_S)$, proving the second piece.

Applying these two bounds to $(-B,-S,-p)$ gives the buyer's central bound for $\mu_B-\sigma_B/3\le p\le\mu_B+\sigma_B/3$ and tail bound for $p>\mu_B+\sigma_B/3$. These prove the fourth and fifth pieces. The third piece is simply $\alpha(p)\le1$. The five cases cover $\R$, regardless of the ordering of the two means.
\end{proof}

By Claim~\ref{cl:decomposition}, we can now integrate the uniform bound in Claim~\ref{cl:trade-profile} over all of $\R$. The seller's tail and central pieces have integrals
\[
 \int_{-\infty}^{\mu_S-\sigma_S/3}
     \frac{2\sigma_S\overline D}{(\mu_S-p)^2}\,dp
 =6\overline D,
 \qquad
 \int_{\mu_S-\sigma_S/3}^{\mu_S+\sigma_S/3}
     \frac{225\overline D}{2\sigma_S}\,dp
 =75\overline D.
\]
The buyer's tail and central interval have the same respective integrals; restricting them to $p>\mu_S+\sigma_S/3$ can only decrease their contribution. The plateau has length $(\mu_B-\sigma_B/3-\mu_S-\sigma_S/3)_+=(m-\sigma/3)_+$. Therefore,
\begin{equation}\label{eq:trade-area}
 G=\int_{\R}\alpha(p)\,dp
 \le\int_{\R}H(p)\,dp
 \le162\overline D+(m-\sigma/3)_+.
\end{equation}
Since $m=\E{B-S}\le G$, subtracting $(G-\sigma/3)_+$ gives
\begin{equation}\label{eq:trade-small-gain}
 \min\{G,\sigma/3\}\le162\overline D.
\end{equation}
Finally, $m<3\sigma$, Cauchy--Schwarz, and the triangle inequality imply
\[
 G\le m_++\E{|(B-\mu_B)-(S-\mu_S)|}
 \le m_++\sigma<4\sigma.
\]
Thus $G\le12\min\{G,\sigma/3\}\le1944\overline D$. If $\overline D=0$, every price has zero lost gain. Otherwise, because $1944<2^{11}$, choose any price with $2^{11}D(p)>1944\overline D$. Then $L(p)\le G\le2^{11}D(p)$.
\end{proof}

\subsection{Proof of Lemma~\ref{lem:trade-lottery}}\label{sec:trade-lottery-proof}

Recall {the quadratic variational expression for} $\Phi$, {given in} \eqref{eq:variational-quadratic-Phi}:
\[
    \Phi(\bm,R,\bz) = \Tr R+(\bz-\bm)^\top R^{-1}(\bz-\bm).
\]
For a random vector $U\in\R^n$ with mean $\bmu$ and covariance $\Sigma$, expanding around the mean and applying Lemma~\ref{lem:envelope} with $\bx=0$ gives, for every $\bm\in\R^n$ and $R\succ0$,
\[
 \E{\Phi(\bm,R,U)}
 \ge2\Psi(\Sigma),
\]
with equality when $\bm = \E{U}$ and $R =\sqrt{\Cov(U)}\succ 0$.

\tradelottery*

\begin{proof}
Fix the current state $(\bm,R)$, context $\bx$, and slack $\gamma\in(0,1]$, and write $\Phi_0(\boldsymbol w)=\Phi(\bm,R,\boldsymbol w)$. If $\bx=0$, there is nothing to prove as the gain from trade is zero at any price, so assume $\bx\ne0$. Let $\mathcal A$ be the set of plans consisting of a common price $p_a$ and two successor states $(\bm_a^{(y)},R_a^{(y)})$ for $y\in\{0,1\}$, with each $R_a^{(y)}$ positive definite and block diagonal.

We recall some useful notation. For $\boldsymbol w=(\boldsymbol s,\boldsymbol b)\in K$, the valuations $S$ and $B$ {are given by $S=\bx^\top\boldsymbol s$ and $B=\bx^\top\boldsymbol b$}. For the price $p_a$ associated with the generic plan $a \in \mathcal A$, we {denote by} $Y_a(\boldsymbol w)$ the indicator of whether $S$ and $B$ {permit trade at} $p_a$, and {by} $\ell_a(\boldsymbol w)$ the corresponding loss {in terms of} the gain from trade:
\[
     Y_a(\boldsymbol w)=\mathbf1\{S\le p_a\le B\},\qquad
 \ell_a(\boldsymbol w)=(B-S)_+(1-Y_a(\boldsymbol w)).
\]
We {denote by} $\Phi_a^+(\boldsymbol w)$ the value of the potential under the realized feedback $Y_a(\boldsymbol w)$ for the generic plan $a \in \mathcal A$ with price $p_a$ and successor states $(\bm_a^{(y)},R_a^{(y)})$ for $y\in\{0,1\}$:
\[
    \Phi_a^+(\boldsymbol w)=\Phi(\bm_a^{(Y_a(\boldsymbol w))},R_a^{(Y_a(\boldsymbol w))},\boldsymbol w).    
\]
Finally, we introduce the function $f_a${, which is central to our proof}:
\[
 f_a(\boldsymbol w)=\ell_a(\boldsymbol w)+2^{10}\Phi_a^+(\boldsymbol w)
                         -2^{10}\Phi_0(\boldsymbol w).
\]
The function $f_a$ is defined in such a way that inequality~\eqref{eq:trade-pointwise} in the statement is {equivalent to the existence of} a finite lottery $Q$ satisfying $\Esub{a\sim Q}{f_a(\boldsymbol w)}\le\gamma$ for every $\boldsymbol w\in K$. 
We first show that, for every hypothetical prior, the {infimum of the expected payoff over plans} is nonpositive, {and then} use minimax and discretization to obtain a single such lottery for all hidden pairs.

\paragraph{A plan for each prior.}
Fix any valuation distribution $\pi\in\mathcal P(K)$ over $K$, where $\mathcal P(K)$ is the set of all possible prior distributions of the hidden pair on $K$. We assume that $S$ and $B$ are induced by sampling $W$ according to $\pi$ and {using }the fixed context $\bx$. For any such distribution, Lemma~\ref{lem:toi} supplies a common price $p$ with $L(p)\le2^{11} D(p)$. {For simplicity, let $Y$ denote the bit $\mathbf1\{S\le p\le B\}$. Write $\pi_y=\mathcal L_\pi(W\mid Y=y)$ whenever $\mathbb P_\pi(Y=y)>0$.} {Here, the latter is the conditional law of $W$ given $Y=y$.} Lemma~\ref{lem:directional-sd-decrease}, applied to both marginals, bounds $D(p)$ by the {decrease in} $\Psi_{\mathrm{tr}}$, so
\[
 \Esub{\pi}{\ell_p(W)}=L(p)
 \le2^{11}\left(\Psi_{\mathrm{tr}}(\pi)
                   -\sum_{y \in \{0,1\}}\mathbb P_\pi(Y=y)\Psi_{\mathrm{tr}}(\pi_y)\right).
\]
Here and below, sums over $y$ omit zero-probability branches. For $j\in\{S,B\}$, let $\bmu_{y,j}$ and $\Sigma_{y,j}$ be the mean and covariance of $W_j$ under $\pi_y$. For $\lambda>0$, define a plan $a_\lambda$ with common price $p$ and successor states
\[
 \bm_{a_\lambda}^{(y)}=(\bmu_{y,S},\bmu_{y,B}),\qquad
 R_{a_\lambda}^{(y)}=\operatorname{diag}\!\left(
 (\Sigma_{y,S}+\lambda I_d)^{\nicefrac 12},
 (\Sigma_{y,B}+\lambda I_d)^{\nicefrac 12}\right).
\]
Choose arbitrary admissible successor states on zero-probability branches.
{We add} the perturbation $\lambda I_d$ to ensure that both $\Sigma_{y,S}+\lambda I_d$ and $\Sigma_{y,B}+\lambda I_d$ are positive definite{, so the inverses of their square roots are well defined.}  Applying the quadratic variational identity \eqref{eq:variational-quadratic-Phi} separately to the two marginals gives
\[
 \Esub{\pi}{\Phi_0(W)}\ge2\Psi_{\mathrm{tr}}(\pi),\qquad
 \lim_{\lambda\to0^+}\Esub{\pi}{\Phi_{a_\lambda}^+(W)}
 =2\sum_{y \in \{0,1\}}\mathbb P_\pi(Y=y)\Psi_{\mathrm{tr}}(\pi_y).
\]
{To compute the limit, diagonalize each marginal covariance: an eigenvalue $\tau\ge0$ contributes $\sqrt{\tau+\lambda}+\tau/\sqrt{\tau+\lambda}$ to the corresponding conditional expected successor potential, which tends to $2\sqrt\tau$ as $\lambda\to 0^+$, including when $\tau=0$.} Combining the preceding bounds yields
\[
 \inf_{a\in\mathcal A}\Esub{\pi}{f_a(W)}
 \le\lim_{\lambda\to0^+}\Esub{\pi}{f_{a_\lambda}(W)}
 \le2^{11}\Psi_{\mathrm{tr}}(\pi)-2^{10}\Esub{\pi}{\Phi_0(W)}\le0.
\]
Since $\pi$ was arbitrary, the ``Bayesian value'' $V_{\mathrm B}$ satisfies
\[
 V_{\mathrm B} = \sup_{\pi\in\mathcal P(K)}\inf_{a\in\mathcal A}\Esub{\pi}{f_a(W)}\le0.
\]
The following claim exchanges the order of choice and provides the regularity needed to discretize the resulting lottery. Its proof involves a {nontrivial} application of {Sion's theorem and is deferred to} the end of the section. We introduce the $\inf \sup$ {value}:
\[
    V_{\mathrm W} = \inf_Q\sup_{\boldsymbol w\in K}\Esub{a\sim Q}{f_a(\boldsymbol w)}.
\]
\begin{restatable}{claim}{sion}\label{cl:trade-minimax}
Allowing arbitrary probability lotteries $Q$ on $\mathcal A$, we have
\[
 \sup_{\pi\in\mathcal P(K)}\inf_{a\in\mathcal A}\Esub{\pi}{f_a(W)}=\inf_Q\sup_{\boldsymbol w\in K}\Esub{a\sim Q}{f_a(\boldsymbol w)}.
\]
Moreover, for every $\eta>0$, there is a lottery $Q$ with $\sup_{\boldsymbol w\in K}\Esub{a\sim Q}{f_a(\boldsymbol w)}<V_{\mathrm{W}}+\eta$ whose price distribution has bounded density and compact support, and whose successor states range over a compact set of positive-definite block-diagonal states.
\end{restatable}

\paragraph{A finite lottery on a finite grid.}
Apply Claim~\ref{cl:trade-minimax} with $\eta=\gamma/3$. Since $V_{\mathrm W}\le0$, it gives a lottery whose payoff is uniformly less than $\gamma/3$, with the regularity stated in the claim. Round its common price to the nearest multiple of $1/n$, retaining both successor states. Compact price support ensures that only finitely many grid points occur. The feedback can change only if the original price lies within $1/n$ of $S$ or $B$. The bounded density makes this event have probability $O(1/n)$ uniformly over $\boldsymbol w\in K$. The branch payoffs are uniformly bounded, so the rounding changes the expected payoff uniformly by $o(1)$.

Next replace the two successor states by representatives of a sufficiently fine finite net in their compact range. Uniform continuity of $\Phi$ {on the product of that range and $K$ allows the additional payoff error to be made uniformly as small as desired}; choosing representatives from the range preserves positive definiteness and block diagonality. Take each of these two errors smaller than $\gamma/3$. The resulting lottery has finite support, satisfies~\eqref{eq:trade-pointwise}, and uses common prices on a finite rational grid.
\end{proof}

\paragraph{Sion's theorem.}
Before proving {\Cref{cl:trade-minimax},} we recall Sion's minimax theorem in its general semicontinuous form and record the affine version used in its proof. {Recall that} $K$ is defined as $\mathbb B_2^d\times\mathbb B_2^d\subset\R^{2d}$ and is endowed with the Euclidean metric, so that it is a {nonempty} compact metric space. Quasiconcavity means that every superlevel set is convex, and quasiconvexity means that every sublevel set is convex.

\begin{theorem}[Sion's minimax theorem~{\cite{Sion58}}]\label{thm:sion-minimax}
Let $M$ and $N$ be nonempty convex subsets of real topological vector spaces, and suppose that at least one of $M$ and $N$ is compact. Let $F:M\times N\to\R$ satisfy the following conditions:
\begin{enumerate}
 \item For every $n\in N$, the function $ F(\cdot,n)$ is upper semicontinuous and quasiconcave on $M$.
 \item For every $m\in M$, the function $F(m,\cdot)$ is lower semicontinuous and quasiconvex on $N$.
\end{enumerate}
Then
\[
 \sup_{m\in M}\inf_{n\in N}F(m,n)
 =\inf_{n\in N}\sup_{m\in M}F(m,n).
\]
\end{theorem}

\begin{corollary}[Separately continuous affine payoffs]\label{cor:sion-affine}
 Let $C(K)$ be the space of continuous real-valued functions on $K$. Let $\mathcal G\subset C(K)$ be nonempty and convex, equipped with the uniform norm.
Equip the space $\mathcal P(K)$ of Borel probability measures on $K$ with the weak topology. Suppose that $F:\mathcal G\times\mathcal P(K)\to\R$ is separately continuous in these topologies and affine in each variable. Then
\[
 \inf_{g\in\mathcal G}\sup_{\pi\in\mathcal P(K)}F(g,\pi)
 =\sup_{\pi\in\mathcal P(K)}\inf_{g\in\mathcal G}F(g,\pi).
\]
\end{corollary}
We now turn to the proof of Claim~\ref{cl:trade-minimax}.
\begin{proof}[Proof of Claim~\ref{cl:trade-minimax}]
We start by proving the ``easy'' direction of the equality, via weak duality. 
For any prior $\pi$ and lottery $Q$, draw $W\sim\pi$ and $a\sim Q$ independently. Then
\[
 \begin{aligned}
 \inf_{a\in\mathcal A}\Esub{\pi}{f_a(W)}
 \le\Esub{a\sim Q}{\Esub{W\sim\pi}{f_a(W)}}=\Esub{W\sim\pi}{\Esub{a\sim Q}{f_a(W)}}
 \le\sup_{\boldsymbol w\in K}\Esub{a\sim Q}{f_a(\boldsymbol w)}.
 \end{aligned}
\]
The first and last inequalities use that an average is at least the infimum and at most the supremum. The exchange of expectations follows from Tonelli's theorem after adding the constant $2^{10}\max_K\Phi_0$, which makes the integrand nonnegative. Taking the supremum over $\pi$ and then the infimum over $Q$ gives $V_{\mathrm B}\le V_{\mathrm W}$.
Both values are finite: $f_a\ge-2^{10}\max_K\Phi_0$, and retaining the current state gives a bounded payoff.

We now move to the other direction. Unfortunately, we cannot {directly invoke} Sion's theorem, as the function $f_a(\boldsymbol w)$ {has} the discontinuities induced by $Y_a(\boldsymbol w)$. To smooth it out, we introduce a random perturbation {of the posted price}. More precisely, we first average the discontinuous payoffs over a small random perturbation, apply minimax to the resulting continuous functions, and then let the perturbation vanish. 

Fix any $0<\varepsilon<\nicefrac 12$ and let $U$ be sampled uniformly at random in $[-\varepsilon\norm{\bx},\varepsilon\norm{\bx}]$ ({recall that} $\bx$ is the {nonzero} context we are given). For any realization $u$ of $U$, we define the perturbed vector $T_u\boldsymbol w$ as follows:
\[
    T_u\boldsymbol w=(1-\varepsilon)\cdot \boldsymbol w+u\cdot \boldsymbol v, \quad \text{where } \boldsymbol v=(\nicefrac{\bx}{\norm{\bx}^2},\nicefrac{\bx}{\norm{\bx}^2}).
\]

{By construction,} the ``perturbed'' values are just the original $S$ and $B$, {scaled by $1-\varepsilon$ and shifted by $u$}:

\[
    \bx^\top \left[(1-\varepsilon)\cdot \boldsymbol s+u\cdot \frac{\bx}{\norm{\bx}^2}\right] = (1-\varepsilon)S + u, \quad \bx^\top \left[(1-\varepsilon)\cdot \boldsymbol b+u\cdot \frac{\bx}{\norm{\bx}^2}\right] = (1-\varepsilon)B + u.
\]

Moreover, $T_u$ {maps $K$ into itself}, i.e., $T_uK\subseteq K$. {To see this, we consider the first $d$ components of $T_u(\boldsymbol w)$; the other case is analogous}:

\[
 \left\lVert(1-\varepsilon)\cdot \boldsymbol s+u\cdot \frac{\bx}{\norm{\bx}^2}\right\rVert
 \le (1-\varepsilon)\norm{\boldsymbol s}+\frac{|u|}{\norm{\bx}}
 \le (1-\varepsilon)+\varepsilon
 =1.
\]

For any base plan $a$, draw $U$ and implement the common price and successor states
\[
 p_{a,U}=\frac{p_a-U}{1-\varepsilon},\qquad
 \bm_{a,U}^{(y)}=\frac{\bm_a^{(y)}-U \boldsymbol v}{1-\varepsilon},\qquad
 R_{a,U}^{(y)}=\frac{R_a^{(y)}}{1-\varepsilon}.
\]
Denote this actual plan by $a_U^\varepsilon$. {By construction,} the original valuations $S$ and $B$ {permit trade at} the perturbed price $p_{a,u}$ if and only if the perturbed valuations {permit trade at} the original price, for any plan $a$ and realization $u$ of $U$:
\[
 \mathbf1\!\left\{S\le\frac{p_a-u}{1-\varepsilon}\le B\right\}
 =\mathbf1\{(1-\varepsilon)S+u\le p_a\le (1-\varepsilon)B+u\}.
\]
{The transformation} $T_u$ is used only in the analysis; the learner implements the displayed price and successor states using its sampled $u$. For either branch $y$, substitution into the quadratic potential gives
\[
 \Phi\!\left(\frac{\bm_a^{(y)}-u \boldsymbol v}{1-\varepsilon},
                  \frac{R_a^{(y)}}{1-\varepsilon},\boldsymbol w\right)
 =\frac1{1-\varepsilon}\Phi(\bm_a^{(y)},R_a^{(y)},T_u\boldsymbol w).
\]
Also, $((1-\varepsilon)B+u-(1-\varepsilon)S-u)_+=(1-\varepsilon)(B-S)_+$, so the lost gain has the same scaling. Combining these identities and subtracting the unchanged current potential gives

\[
 g_a^\varepsilon(\boldsymbol w)
 =\Esub{U}{f_{a_U^\varepsilon}(\boldsymbol w)}
 =\frac1{1-\varepsilon}\Esub{U}{ f_a(T_U\boldsymbol w)}
   +2^{10}\Esub{U}{\frac{\Phi_0(T_U\boldsymbol w)}{1-\varepsilon}-\Phi_0(\boldsymbol w)}.
\]

    The crucial fact is that now, for each fixed $a$, the averaged payoff $g_a^\varepsilon$ is continuous in $\boldsymbol w$. To see this, consider any sequence $\boldsymbol w_n\to\boldsymbol w$. Except at the two noise values $u=p_a-(1-\varepsilon)S$ and $u=p_a-(1-\varepsilon)B$, the feedback at $\boldsymbol w_n$ eventually agrees with that at $\boldsymbol w$, and the corresponding branch payoff converges. These exceptional values have probability zero under $U$. Both branch payoffs are bounded on the compact set of hidden pairs ($\boldsymbol{w} \in K$) and noise values ($|u|\le\varepsilon\norm{\bx}$ ), so dominated convergence applies, including when $S=B$.

Moreover, the absolute value of the difference between $\Phi_0$ and its scaled version converges uniformly to $0$:
\[
 \delta_\varepsilon
 =2^{10}\sup_{\substack{\boldsymbol w\in K\\ |u|\le\varepsilon\norm{\bx}}}
 \left|\frac{\Phi_0(T_u\boldsymbol w)}{1-\varepsilon}-\Phi_0(\boldsymbol w)\right|
 \longrightarrow0.
\]
Indeed, $\norm{T_u\boldsymbol w-\boldsymbol w}\le2\sqrt2\,\varepsilon$, and the current potential $\Phi_0$ is uniformly continuous on $K$. In particular, the correction is controlled independently of the base plan $a$.

For a prior $\pi$, let $\pi_\varepsilon$ be the law of $T_UW$ when $W\sim\pi$ and $U$ is independent. Since $T_uK\subseteq K$, this is another prior in $\mathcal P(K)$. Averaging the payoff identity and using the definition of $V_{\mathrm B}$ yields
\[
 \inf_a\Esub{\pi} {g_a^\varepsilon(W)}
 \le\frac1{1-\varepsilon}\inf_a\Esub{W'\sim\pi_\varepsilon}{f_a(W')}+\delta_\varepsilon
 \le\frac{V_{\mathrm B}}{1-\varepsilon}+\delta_\varepsilon.
\]

We are now ready to {apply} the version of Sion's theorem presented in \Cref{cor:sion-affine}. To this end, we consider the space $C(K)$ of {continuous} functions from $K$ to $\R$, equipped with the uniform norm; in particular, we focus on $\mathcal G_\varepsilon\subset C(K)$, which is defined as the convex hull of the functions $g_a^\varepsilon$.
By definition, every $g\in\mathcal G_\varepsilon$ has the form
\[
 g=\sum_{i=1}^N\alpha_i g_{a_i}^\varepsilon,
 \qquad \alpha_i\ge0,\qquad \sum_{i=1}^N\alpha_i=1,
\]
for finitely many base plans $a_1,\ldots,a_N$. To implement $g$, first sample an index $i$ with probabilities $(\alpha_i)_{i=1}^N$, then independently sample $U$ as above and execute the plan $(a_i)_U^\varepsilon$. The resulting lottery $Q_g$ on $\mathcal A$ satisfies, for every hidden pair,
\[
 \Esub{a\sim Q_g}{f_a(\boldsymbol w)}
 =\sum_{i=1}^N\alpha_i\Esub{U}{f_{(a_i)_U^\varepsilon}(\boldsymbol w)}
 =g(\boldsymbol w).
\]
Here the mixture uses finitely many base plans, but $Q_g$ itself need not have finite support because $U$ has a continuous distribution. Such lotteries are allowed in the definition of $V_{\mathrm W}$; finite support will be obtained in the subsequent discretization step. Thus minimizing over $g\in\mathcal G_\varepsilon$ restricts the admissible lotteries and can only increase the infimum defining $V_{\mathrm W}$.

Define the payoff pairing
\[
 F(g,\pi)=\int_K g(\boldsymbol w)\,d\pi(\boldsymbol w).
\]
{Recall that} $\mathcal G_\varepsilon$ is equipped with the uniform norm $\|g\|_\infty=\max_{\boldsymbol w\in K}|g(\boldsymbol w)|$, while $\mathcal P(K)$ {is equipped }with the topology of weak convergence of probability measures. Since $K$ is compact, $\mathcal P(K)$ is compact in this topology; it is also convex. The set $\mathcal G_\varepsilon$ is convex by construction, and $F$ is affine in each argument separately.

{
Explicitly, for $t\in[0,1]$, $g,h\in\mathcal G_\varepsilon$, and $\pi,\pi_1,\pi_2\in\mathcal P(K)$, linearity of integration gives
\[
 \begin{aligned}
 F(tg+(1-t)h,\pi)
 &=tF(g,\pi)+(1-t)F(h,\pi),\\
 F(g,t\pi_1+(1-t)\pi_2)
 &=tF(g,\pi_1)+(1-t)F(g,\pi_2).
 \end{aligned}
\]
These are exactly the affine identities required in Corollary~\ref{cor:sion-affine}. We next check separate continuity, meaning continuity in either variable with the other held fixed.
}

For a fixed prior $\pi$ and any $g,h\in\mathcal G_\varepsilon$,
\[
 |F(g,\pi)-F(h,\pi)|
 \le\int_K|g-h|\,d\pi
 \le\|g-h\|_\infty.
\]
For a fixed $g\in\mathcal G_\varepsilon$, weak convergence $\pi_n\to\pi$ gives
\[
 F(g,\pi_n)=\int_Kg\,d\pi_n
 \longrightarrow\int_Kg\,d\pi=F(g,\pi),
\]
because $g$ is continuous and bounded on $K$. This is precisely where continuity of the smoothed payoffs is used.

{
We can therefore apply Corollary~\ref{cor:sion-affine} with $\mathcal G=\mathcal G_\varepsilon$ and the pairing $F(g,\pi)=\int_Kg\,d\pi$. The hidden-pair set $K$ is a nonempty compact metric space, $\mathcal G_\varepsilon\subset C(K)$ is nonempty and convex with the uniform norm, and $\mathcal P(K)$ carries the weak topology. The two preceding calculations verify separate affinity and continuity.  We obtain
}
\[
 \inf_{g\in\mathcal G_\varepsilon}\sup_{\pi\in\mathcal P(K)}F(g,\pi)
 =\sup_{\pi\in\mathcal P(K)}\inf_{g\in\mathcal G_\varepsilon}F(g,\pi).
\]

For each fixed $g$, an average under any prior is at most $\max_K g$. Conversely, $g$ attains this maximum at some $\boldsymbol w_g\in K$, and the point mass $\delta_{\boldsymbol w_g}$ attains the same value in the prior formulation. Hence
\[
 \sup_{\pi\in\mathcal P(K)}F(g,\pi)
 =F(g,\delta_{\boldsymbol w_g})
 =\max_{\boldsymbol w\in K}g(\boldsymbol w).
\]
For each fixed prior $\pi$, linearity gives, for every finite mixture,
\[
 F\!\left(\sum_{i=1}^N\alpha_i g_{a_i}^\varepsilon,\pi\right)
 =\sum_{i=1}^N\alpha_i\Esub{\pi}{g_{a_i}^\varepsilon(W)}
 \ge\inf_{a\in\mathcal A}\Esub{\pi} {g_a^\varepsilon(W)}.
\]
Since each individual $g_a^\varepsilon$ also belongs to the convex hull, the reverse inequality holds after taking the infimum over $g$. Thus
\[
 \inf_{g\in\mathcal G_\varepsilon}F(g,\pi)
 =\inf_{a\in\mathcal A}\Esub{\pi}{g_a^\varepsilon(W)}.
\]
Combining the implementation of $g$, the point-mass identity, Sion's exchange, and the convex-hull identity, in that order, gives
\[
 \begin{aligned}
 V_{\mathrm W}
 &\le\inf_{g\in\mathcal G_\varepsilon}\max_{\boldsymbol w\in K}g(\boldsymbol w)\\
 &=\inf_{g\in\mathcal G_\varepsilon}\sup_{\pi\in\mathcal P(K)}F(g,\pi)\\
 &=\sup_{\pi\in\mathcal P(K)}\inf_{g\in\mathcal G_\varepsilon}F(g,\pi)\\
 &=\sup_{\pi\in\mathcal P(K)}\inf_{a\in\mathcal A} \Esub{\pi}{g_a^\varepsilon(W)}
 \le\frac{V_{\mathrm B}}{1-\varepsilon}+\delta_\varepsilon.
 \end{aligned}
\]
The last inequality is the bound proved above for every prior $\pi$, whose right-hand side does not depend on $\pi$. This estimate holds for every $0<\varepsilon<1/2$. Since $V_{\mathrm B}$ is finite, letting $\varepsilon\to 0^+$ gives $V_{\mathrm W}\le V_{\mathrm B}$. Together with weak duality, this proves equality: $V_{\mathrm W}=V_{\mathrm B}$. The limit is taken only in this numerical bound, so no convergence of the implementing lotteries is needed.

For the additional assertion, choose $\varepsilon$ small enough and then an element of $\mathcal G_\varepsilon$ whose maximum is less than $V_{\mathrm{B}}+\eta$; no attainment of the infimum is required. This element is implemented by a finite mixture of the smoothed plans $a_U^\varepsilon$. Each common price is uniform on a bounded interval of positive length, and each successor state is a continuous function of $U$ on a compact interval. The finite mixture therefore has bounded price density, compact price support, and a compact range of positive-definite block-diagonal successor states.

More explicitly, every price interval has length $2\varepsilon\norm{\bx}/(1-\varepsilon)$, so the mixture's density is bounded by $(1-\varepsilon)/(2\varepsilon\norm{\bx})$. The choice of $\varepsilon>0$ and the finite mixture is fixed before the subsequent grid approximation.
\end{proof}

\begin{proof}[{Proof of Corollary~\ref{cor:sion-affine}}]
Since $K$ is compact, $\mathcal P(K)$ is a compact convex subset of the space of finite signed Borel measures equipped with the weak topology. Apply Theorem~\ref{thm:sion-minimax} with $M=\mathcal P(K)$, $N=\mathcal G$, and the transposed payoff $\widetilde F(\pi,g)=F(g,\pi)$. For fixed $g$, continuity in $\pi$ implies upper semicontinuity, and affinity implies concavity and hence quasiconcavity. For fixed $\pi$, continuity in $g$ implies lower semicontinuity, and affinity implies convexity and hence quasiconvexity. Thus all the hypotheses hold, with compactness supplied by the prior space $\mathcal P(K)$, and the theorem gives the stated equality.
\end{proof}
\section{Lower Bounds}\label{app:lower-bound}\label{sec:trade-lower-bound}

In this section, we present standard lower bounds for both contextual search and contextual bilateral trade.
First, we recall the standard coordinate-wise lower-bound argument for contextual search \cite{LPS21}.

\begin{proposition}[Lower bound for contextual search]\label{prop:lower-bound}
For every $d\ge1$, every $T\ge d$, and every possibly randomized algorithm, there exists a fixed target $\bz\in \Rbb^d$ with $\norm{\bz}=1$ and a deterministic sequence of context vectors of norm at most one such that
\[
 \sum_{t=1}^T\E{\ell_t}\ge\sqrt d.
\]
\end{proposition}
\begin{proof}
We consider an adversary that chooses the coordinates of $\bz$ independently and uniformly from $\{-1/\sqrt d,1/\sqrt d\}$, so that $||\bz||=1$. For the first $d$ rounds, the adversary lets $\bx_t$ be the $t$-th basis vector, and sets $\bx_t=0$ afterward. At round $t\leq d$, $z_t = \bx_t^\top \bz$ is independent of any previous observation {made by} the algorithm, and thus of the posted price $P_t${. Therefore,} 
\[
 \E{\ell_t\mid P_t}
 =\frac12\left|P_t-\frac1{\sqrt d}\right|
  +\frac12\left|P_t+\frac1{\sqrt d}\right|
 \ge\frac1{\sqrt d}.
\]
Summing the loss over $t\in \{1,\dots d\}$ provides the desired lower bound. 
\end{proof}

We now turn to the matching lower bound for contextual bilateral trade. The coordinate construction in \cite[Proposition~4.3]{CFGLPR26}, rescaled to the Euclidean unit ball, gives the following bound. {Here, we allow} the learner to post two distinct prices, as long as budget balance is enforced, i.e., the {price $p_t$ offered to the seller} is at most {the price $q_t$ offered to the buyer}.

\begin{proposition}[Lower bound for contextual bilateral trade]\label{prop:trade-lb}
For every $d\ge1$, every $T\ge d$, and every possibly randomized learner satisfying $p_t\le q_t$, there are a fixed pair $(\boldsymbol s,\boldsymbol b)\in K$ and deterministic contexts of norm at most one for which
\[
 \sum_{t=1}^T\E{\ell_t}\ge\frac{\sqrt d}{6}.
\]
The same lower bound holds in the Bayesian setting for the prior whose coordinates are independent and for which each pair $(\sqrt d\,\bs_j,\sqrt d\,\bb_j)$ is uniform on $\{(0,1/3),(2/3,1)\}$.
\end{proposition}
\begin{proof}
This prior is supported on $K$, since $\norm{\bs}^2\le4/9$ and $\norm{\bb}^2\le1$. Use the $t$-th coordinate context on rounds $1\le t\le d$, and zero contexts afterwards. Each round has gain $1/(3\sqrt d)$, but its trading interval is equally likely to be $[0,1/(3\sqrt d)]$ or $[2/(3\sqrt d),1/\sqrt d]$. Conditional on the learner's random seed and previous coordinates, the current interval remains independent and uniform. The nonempty interval $[p_t,q_t]$ can lie inside at most one of these two disjoint intervals, so trade occurs with probability at most $1/2$. Each of the first $d$ rounds therefore loses at least $1/(6\sqrt d)$ in expectation. Summing gives at least $\sqrt d/6$, and later losses are nonnegative. Averaging over this finite prior selects a fixed pair with at least the same expected loss, even when the learner is given the prior.
\end{proof}

Both marginal covariances of this prior are $I_d/(9d)$, so $\Psi_{\mathrm{tr}}(\pi)=2\sqrt d/3$. Thus the Bayesian lower bound is $\Psi_{\mathrm{tr}}(\pi)/4$, and the dependence on the trace-square-root potential cannot be improved by more than a constant factor uniformly over priors. The dimension dependence in Theorems~\ref{thm:trade-bayes} and~\ref{thm:trade-direct} is consequently optimal.

\end{document}